\documentclass{aa}
\usepackage[varg]{txfonts}
\usepackage{natbib,twoopt}
\usepackage[breaklinks=true]{hyperref} 
\bibpunct{(}{)}{;}{a}{}{,} 
\usepackage{graphicx}
\usepackage{subcaption} 

\begin{document}

\title{The stellar content of the ROSAT all-sky survey}
\author{S. Freund\inst{1}\inst{2} \and S. Czesla\inst{1}\inst{3} \and J. Robrade\inst{1} \and P.C. Schneider\inst{1} \and J.H.M.M. Schmitt\inst{1}}
\institute{Hamburger Sternwarte, Universit\"at Hamburg, 21029 Hamburg, Germany
\and
Max-Planck-Institut für extraterrestrische Physik, Gießenbachstraße 1, 85748 Garching Germany
\email{sfreund@mpe.mpg.de}
\and
Th\"uringer Landessternwarte Tautenburg, Sternwarte 5, D-07778 Tautenburg, Germany }
\abstract{}
{We present and apply a method to identify the stellar content of the \textit{ROSAT} all-sky survey (RASS).}
{
We performed a crossmatch between the RASS sources and stellar candidates selected from Gaia Early Data Release 3 (EDR3) and estimated stellar probabilities for every RASS source from the geometric properties of the match and additional properties, namely the X-ray to G-band flux ratio and the counterpart distances.}
{A comparison with preliminary detections from the first \textit{eROSITA} all-sky survey (eRASS1) show that the positional offsets of the RASS sources are larger than expected from the uncertainties given in the RASS catalog. From the RASS sources with reliable positional uncertainties, we identify 28\,630 (24.9~\%) sources as stellar; this is the largest sample of stellar X-ray sources to date. Directly from the stellar probabilities, we estimate the completeness and reliability of the sample to be about 93~\% and confirm this value by comparing it to the identification of randomly shifted RASS sources, preliminary stellar eRASS1 identifications, and results from a previous identification of RASS sources. Our stellar RASS sources contain sources of all spectral types and luminosity classes. According to their position in the color-magnitude diagram, many stellar RASS sources are young stars with ages of a few $10^7$~yr or binaries. When plotting the X-ray to bolometric flux ratio as a function of the color, the onset of convection and the saturation limit are clearly visible. We note that later-type stars reach continuously higher $F_\mathrm{X}/F_\mathrm{bol}$ values, which is probably due to more frequent flaring. The color distribution of the stellar RASS sources clearly differs from the unrelated background sources. We present the three-dimensional distribution of the stellar RASS sources that shows a clear increase in the source density near known stellar clusters.}
{}
\maketitle

\section{Introduction}
\label{sec: introduction}
The X-ray observatory \textit{ROSAT} \citep{truemper84}, which launched in June 1990, performed the \textit{ROSAT} all-sky survey (RASS) between August 1990 and January 1991 with the 
positional-sensitive proportional counter (PSPC); the resulting catalogs were published by \cite{vog99}, and later
using improved processing and analysis algorithms by \cite{RASS-catalog}.   In the past,
the X-ray properties of RASS-detected stellar sources were investigated by selecting specific subsamples, for example
RS CVn systems \citep{dempsey93}, OB-type stars \citep{ber97}, or volume-limited or flux-limited samples of late-type and/or giant stars 
\citep{schmitt95, huensch96, schmitt97, huensch98a, huensch98b, huensch99, schmitt04}. Stellar samples from \textit{Chandra} pointings, RASS sources that overlap with the Sloan Digital Sky Survey (SDSS), and the \textit{XMM-Newton} slew survey (XMMSL) have been provided by \citet{covey08}, \citet{agueros09}, and \citet{freund18}.

The RASS catalog has also been cross-correlated with optical and IR all-sky catalogs; for example, identifications based on
the \textit{Tycho} and \textit{Hipparcos} catalogs were presented by \citet{guillout99}, while \citet{haakonsen09} identified bright RASS sources 
with the \textit{2MASS} Point Source Catalog, however,  without providing a classification of the source types. \citet{salvato18} present an identification of high latitude RASS sources using the \textit{AllWISE} catalog with a special emphasis on the identification of active galactic nuclei (AGN); they also provide stellar identifications classified by a relation between X-ray and \textit{AllWISE} fluxes. Because \citet{salvato18} focus on extragalactic objects, they exclude sources near the Galactic plane or the Large and Small Magellanic Clouds where confusion with Galactic foreground objects is an issue. Since we are specifically interested in these Galactic foreground objects and as a significant number of stellar sources lie, in fact, in the Galactic plane, the exclusion of this sky region is problematic from a stellar point of view. 
In summary, the full stellar content of RASS has so far never been identified and new data such as the \textit{Gaia} survey put us in the position to accomplish this task.

The optical brightness of counterparts to stellar X-ray sources at a specific X-ray luminosity level
can substantially vary due to different X-ray production mechanisms. The X-ray emission of early-type stars is thought to be
generated through instabilities in their radiatively driven stellar winds and reach a typical fractional contribution of the X-ray to the total energy 
output of $L_\mathrm{X}/L_\mathrm{bol} \approx 10^{-7}$ \citep{pal81,ber97}, thus these OB-type stars are relatively X-ray faint
when compared to their optical brightness.
On the other hand, for late-type stars with a convective envelope, that is to say with stellar masses between 0.08 and 
1.85 $M_\odot$ (spectral types M to mid A), the X-ray emission is thought to be produced by some magnetic field related heating 
mechanism in a hot corona. The X-ray activity strongly correlates with stellar rotation and age \citep{wilson63,skumanich72,preibisch05}, and hence 
the X-ray luminosities substantially differ even for stars of the same mass. 

The lowest levels of X-ray activity are observed at $L_\mathrm{X}/L_\mathrm{bol} \approx 10^{-8}$ \citep[see][and references therein]{gud04,testa15} 
for old slowly rotating stars, while the X-ray emission saturates at $L_\mathrm{X}/L_\mathrm{bol} \approx 10^{-3}$ \citep{vil84,wright11} 
for very young and fast rotating stars. The physical causes of this saturation phenomenon are still under debate. Possible explanations for the saturation include the saturation of the dynamo efficiency \citep{gilman83}, the complete coverage of the stellar surface with active regions \citep{vil84}, or a change of the underlying dynamo mechanism \citep{barnes03}. Also, evolved stars show X-ray emission with -- in some cases -- very high X-ray luminosities, especially when 
the rotation period is preserved by a tidal interaction with a binary component as in RS CVn and related systems \citep{wal78,dempsey93}. 
Very little X-ray emission is found for red giants beyond the so-called dividing line \citep{linsky79, huensch96}. Furthermore, 
the observed X-ray luminosity of individual objects can increase over timescales of minutes to hours and, in some cases,
by orders of magnitude when observed during a flare. On timescales of years, the X-ray emission of late-type stars may 
show modulations related to activity cycles similar to the solar cycle \citep{Hempelmann_2003,Robrade_2012}. 

Very importantly, as a consequence of the saturation limit, the stellar counterparts in X-ray surveys with a given limiting X-ray flux will also be limited in their
optical flux, and therefore a complete census of the stellar RASS source requires a homogeneous counterpart catalog of sufficient depth. 
Furthermore, additional properties of the counterparts ought to be available to distinguish between stellar and nonstellar counterparts and between likely 
identifications and unrelated background sources within the relatively large error circles of the RASS positions. 
Such a catalog is provided by the \textit{Gaia} mission \citep{GaiaMission} which contains, in its current version \textit{Gaia} EDR3 \citep{GaiaEDR3}, highly accurate positions, proper motions, 
magnitudes, colors, and parallaxes for more than 1.4 billion sources down to the 21\textsuperscript{st} magnitude. 
Specifically, given a typical RASS sensitivity of 1.5 $\times$ 10$^{-13}$~erg\,s$^{-1}$\,cm$^{-2}$, we expect most counterparts to be brighter than ({\it Gaia}) magnitude G $\approx$ 15
and be located within a distance of a few hundred parsecs.    Therefore, we assume that all stellar counterparts of RASS X-ray sources are, first, contained in \textit{Gaia} EDR3
and, second, have a parallax measurement.

Thus, the \textit{Gaia} EDR3 catalog puts us into the position to identify -- for the first time --  the whole stellar content of RASS as a 
sample of stellar sources only limited by their X-ray fluxes, but not biased by any preselection of specific sources of interest. In this paper, we therefore present an identification method for stellar X-ray sources and provide a definitive identification of the stellar content of RASS. The \textit{Gaia} parallaxes allow -- again for the first time -- one to accurately estimate X-ray luminosities of all stellar RASS sources, to accurately place them in the Hertzsprung-Russell diagram, and thence evaluate their three-dimensional spatial distribution.

We structured our paper as follows: in Sect.~\ref{sec: catalogs} we present the input and matching catalogs namely the RASS catalog and \textit{Gaia} EDR3; furthermore, we introduce the \textit{eROSITA} all-sky survey (eRASS) which we use for verification. Next, we discuss the positional 
uncertainties of the RASS sources, define our sample of stellar candidates, and describe our identification procedure in Sect.~\ref{sec: matching procedure}. We present our 
results in Sect.~\ref{sec: results} and compare our stellar identifications with the first \textit{eROSITA} all-sky survey (eRASS1) and \citet{salvato18}. In Sect.~\ref{sec: properties of stellar RASS sources} we discuss the properties 
of the obtained sample of stellar RASS sources and we draw our conclusions in Sect.~\ref{sec: conclusion}.

\section{Input and matching catalogs}
\label{sec: catalogs}
\subsection{Second \textit{ROSAT} all-sky survey (2RXS) source catalog}
\label{sec: RASS catalog}
During its all-sky survey, the \textit{ROSAT} satellite scanned the sky along great circles over the ecliptic poles, resulting in an exposure time varying between about 400~s and 40000~s for most parts of the sky \citep{vog99}. From the reduction of these data, \citet{RASS-catalog} created the second \textit{ROSAT} all-sky survey (2RXS) source catalog (hereafter: RASS catalog) and provide a detailed description of the content of the catalog.  

For our work the following properties of the RASS catalog are of particular importance: the RASS catalog contains about 135\,000 X-ray sources in the $0.1 - 2.4$~keV energy band down to a detection likelihood of 6.5, yet, depending on detection likelihood, up to 30~\% of the sources are expected to be spurious detections. Due to the observing strategy, the exposure time strongly varies with the ecliptic latitude, and hence the RASS catalog has no uniform detection limit. We adopt the count rate to flux conversion of \citet{schmitt95} throughout this paper, leading to a detection limit ranging from $\sim 3\cdot10^{-14}$ -- $5\cdot10^{-13}\,$erg\,s$^{-1}$\,cm$^{-2}$ for 99~\% of the RASS sources with a mean detection limit of approximately $1.5\cdot10^{-13}\,$erg\,s$^{-1}$\,cm$^{-2}$. \citet{RASS-catalog} provide statistical uncertainties of the RASS source positions that partly depend on the number of detected counts and have a mean dispersion of 11.9~arcsec;  we discuss the positional accuracy of the RASS sources in detail in Sect.~\ref{sec: Positional uncertainties of the RASS sources}.

\subsection{\textit{Gaia} EDR3}
\label{sec: Gaia EDR3 catalog}
The current version of the \textit{Gaia} catalog, \textit{Gaia} EDR3, is based on 34 months of data collected between July 2014 and May 2017. In the following we provide a description of the properties and limitations of \textit{Gaia} EDR3 that are most important in the context of this paper, and we refer readers to \citet{GaiaEDR3} and \citet{fabricius21} for a detailed description. 

\textit{Gaia} EDR3 provides highly accurate positions with typically submilliarcsecond uncertainties at epoch J2016. More than 1.4 of the 1.8 billion \textit{Gaia} EDR3 sources further contain proper motions and parallaxes; their typical uncertainties are smaller than 0.5~mas and strongly decrease with increasing brightness of the source. Most of the sources without proper motion and a parallax are very faint, especially for sources fainter than about $G=21$~mag only the position is provided, but also 1.5~\% of the sources brighter than $G=19$~mag miss parallaxes and proper motions in many cases due to problems with a close neighbor. Furthermore, some parallaxes are unreliable, specifically 1.6~\% of the sources with \texttt{parallax\_over\_error} $>5$ are expected to be spurious, the fraction strongly decreasing with source magnitude. The \textit{Gaia} EDR3 catalog also contains broadband photometry in the G, BP, and RP band. For about 5.4 million sources, the G-band magnitude is missing and nearly 300 million sources do not have a BP or RP magnitudes 
due to processing problems.

\textit{Gaia} EDR3 is essentially complete between $G = 12$ and 17~mag; however, the completeness is reduced for sources brighter than $G=7$~mag and 20~\% of the stars brighter than 3~mag are missing. Furthermore, the catalog is incomplete in regions with source densities above $6\times 10^5$ stars deg$^{-2}$. Binaries are resolved for separations above 2~arcsec and the completeness decreases rapidly for separations below 0.7~arcsec. Even for resolved binaries, the parallax or the magnitude in one band might be missing due to processing problems.

\subsection{eROSITA all-sky survey}
\label{sec: eRASS catalog}
In July 2019, the extended ROentgen Survey with an Imaging Telescope Array (\textit{eROSITA}) instrument was launched onboard the 
Spectrum-Roentgen-Gamma (SRG) mission and started its four-year-long all-sky survey in December 2019;  a detailed description of the 
\textit{eROSITA} hardware, mission, and in-orbit performance is presented by \citet{predehl21}. 
In the summer of 2020, eRASS1 was completed and the preliminary data in the western half of 
the sky with Galactic longitudes $l>180^\circ$ are processed by the eSASS pipeline (currently version 946) \citep{brunner21}. The scanning law of eRASS1 is similar to RASS with varying exposure times between the ecliptic poles and plane. On average, a detection limit of about $5\cdot10^{-14}\,$erg\,s$^{-1}$\,cm$^{-2}$ is reached for stellar sources, and the mean positional uncertainty of the eRASS1 sources is $\sigma_r = 3.4$~arcsec. However, the accuracy depends on the number of detected counts so that the mean dispersion of the positional uncertainty for eRASS1 sources with RASS identification is about 1.8~arcsec.

With a flux measurement obtained 30 years ago, the RASS survey provides a very important point of
comparison to many sources detected by \textit{eROSITA}.\ Furthermore,
since the positional accuracy of the eRASS1 sources is much better than that of the RASS sources, the eRASS1 sources are ideally suited to check and verify the RASS positions and stellar identifications.

\section{Matching procedure}
\label{sec: matching procedure}

We seek to identify the full stellar content of RASS and, more specifically, the coronal X-ray emitters detected in RASS. 
In this context we also include OB-type stars when we speak of coronal sources even though the X-ray emission in these sources
is produced by stellar winds, yet the X-ray properties are rather similar to proper coronal sources.

In our first paper, we identified the stellar content of the \textit{XMM-Newton} slew survey (XMMSL) \citep{freund18}, applying only 
the angular separation as a matching criterion. This is suitable for XMMSL because the positions of the XMMSL sources are quite
accurate and the density of the candidate stellar counterparts is quite small. Therefore, most XMMSL sources  only have one 
plausible stellar counterpart. However, for the RASS sources we often find several stellar candidates in the search region 
because the RASS positions are less accurate and the RASS catalog is more sensitive than XMMSL and thus, the stellar counterparts 
are expected to be fainter. Therefore, additional source properties have to be considered to find the correct identification.

In Sect.~\ref{sec: Positional uncertainties of the RASS sources} we discuss the positional uncertainties of the RASS sources and 
in Sect.~\ref{sec: Candidate counterparts and source densities} we define our candidate stellar counterparts and estimate their 
source density. We describe the estimation of the matching probabilities adopting only geometric properties in 
Sect.~\ref{sec: matching probability}, and we consider additional properties by applying a Bayes map in Sect.~\ref{sec: considering additional properties}.

\subsection{Positional uncertainties of the RASS sources}
\label{sec: Positional uncertainties of the RASS sources}
\begin{figure}[t]
        \includegraphics[width=\hsize]{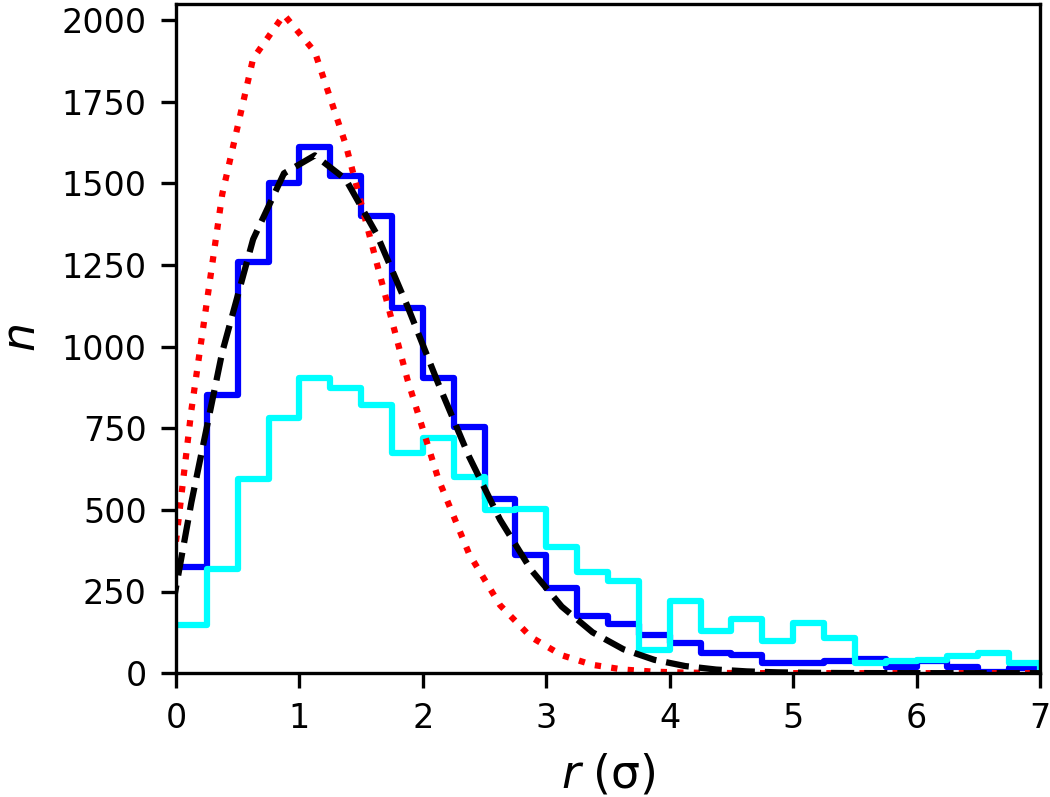}
        \caption{Nearest neighbor distribution for the RASS and the stellar eRASS1 sources as a function of the positional accuracy given in the RASS catalog. The blue and cyan solid histograms show the nearest neighbor distributions for the main and supplement RASS sources (see text for details). Both histograms are normed to the same number of sources. The red and black dashed line show the Rayleigh distribution
        (Equation~\ref{equ: Rayleigh distribution}) and the best fit of the blue histogram  applying Equation~\ref{equ: positional uncertainty}, respectively.}
        \label{fig: RASS stellar eRASS nearest neighbor}
\end{figure}

We tested the positional offsets of the RASS sources by crossmatching them with proper motion corrected eRASS1 sources identified as stellar by a preliminary method (see Sect.~\ref{sec: comparison stellar eRASS1}), the positional uncertainties of which are small compared to that of the RASS sources. 
In Fig.~\ref{fig: RASS stellar eRASS nearest neighbor} we show the angular separations between the RASS sources and the nearest stellar eRASS1 counterparts scaling the separations with the positional uncertainties given in the RASS catalog. To take the number of spurious associations into account, we subtracted the distribution of the separations of the eRASS1 sources 
in Fig.~\ref{fig: RASS stellar eRASS nearest neighbor} to randomly shifted RASS sources from the distribution of the real RASS sources. The offsets are expected to be described by the Rayleigh distribution through 
\begin{equation}
n(r) = \frac{r}{\sigma_\mathrm{RASS}^2+\sigma_\mathrm{eRASS1}^2} e^{-\frac{r^2}{2\left(\sigma_\mathrm{RASS}^2+\sigma_\mathrm{eRASS1}^2\right)}},
\label{equ: Rayleigh distribution}
\end{equation} 
where $\sigma_\mathrm{RASS}$ and $\sigma_\mathrm{eRASS1}$ are the positional uncertainties given in the RASS and eRASS1 catalogs and $r$ is the angular separation between the RASS and eRASS1 sources. However, as shown in Fig.~\ref{fig: RASS stellar eRASS nearest neighbor}, the Rayleigh distribution does not describe the offsets of the RASS positions well, and hence we fit the RASS positional uncertainties in arcsec by applying the ansatz
\begin{equation}
\sigma_r = s\cdot\sqrt{\frac{(\texttt{XERR}\cdot 45)^2 + (\texttt{YERR}\cdot 45)^2}{2} + \sigma_{sys}^2}, 
\label{equ: positional uncertainty}
\end{equation} 
where \texttt{XERR} and \texttt{YERR} are the uncertainties given in the RASS catalog in detector coordinates, converted to arcsec by a factor of 45, $\sigma_{sys}$ is the systematic uncertainty, and $s$ is a scaling factor. 

We find that the nearest neighbor distribution can be fitted well by Equation~\ref{equ: positional uncertainty} (see Fig.~\ref{fig: RASS stellar eRASS nearest neighbor}); however, a fraction of identifications still have a larger positional offset than expected from a Gaussian distribution and this fraction gradually increases with the positional uncertainty given in the RASS catalog. To restrict the fraction to about 5~\%, we discuss in the following only the 115\,000 RASS sources with the best positional accuracies, which we refer to as the main RASS catalog. For these sources, the nearest neighbor distribution is best fitted by a systematic uncertainty of ${\sigma_{sys}=3}$~arcsec and a scaling factor of $s=1.22$. Hence, the main RASS catalog only contains sources with a positional accuracy of $\sigma_r < 20.4$~arcsec. We provide the identifications of the RASS sources with larger positional uncertainties in a supplementary catalog, but we note that our matching procedure is less reliable for these sources.

\subsection{Candidate counterparts and source densities}
\label{sec: Candidate counterparts and source densities}
At the mean RASS detection limit, a high luminous stellar X-ray source with $L_X=10^{31}$~erg\,s$^{-1}$ can be detected up to a distance of $\sim$750~pc. Since \textit{Gaia} EDR3 provides accurate parallaxes out to much larger distances, we can differentiate between plausible stellar and extragalactic counterparts by only considering \textit{Gaia} EDR3 sources with a parallax significance $>3\sigma$; however, the exact value of the considered parallax value has a minor influence on our results. 

\begin{figure}[t]
        \resizebox{\hsize}{!}{\includegraphics{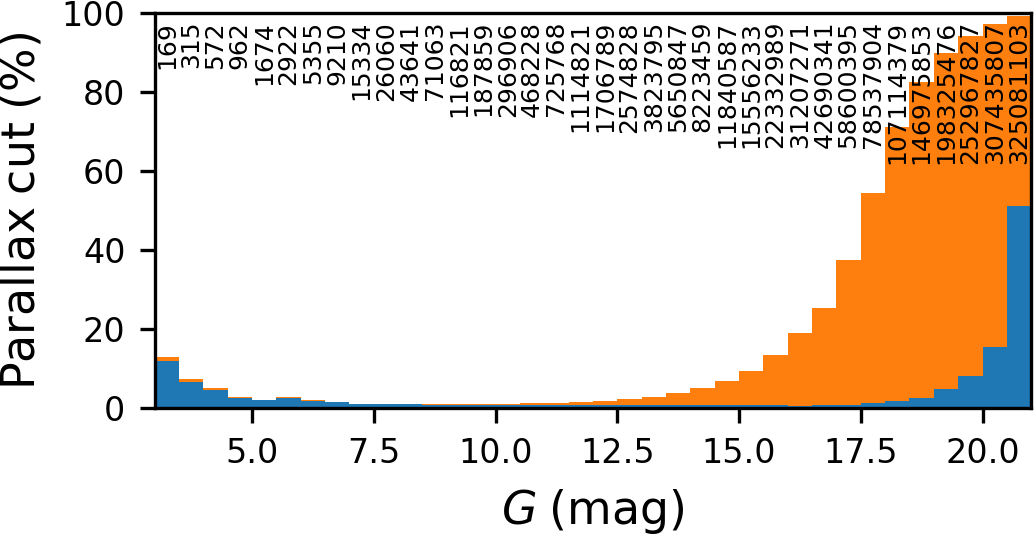}}
        \caption{Fraction of the \textit{Gaia} EDR3 sources that are filtered out by the parallax cutoff as a function of the source magnitude. The blue bars indicate the fraction of sources without a parallax in \textit{Gaia} EDR3, while the orange bars represent the fraction of sources for which the accuracy of the parallax is less than $3\sigma$. The values given at the top indicate the total number of \textit{Gaia} EDR3 sources in the magnitude bin. }
        \label{fig: plx cut}
\end{figure}

In Fig.~\ref{fig: plx cut} we show the fraction of the sources that are filtered out by this parallax cutoff as a function of the source magnitude. Thus, the fraction of \textit{Gaia} EDR3 sources with a low parallax significance that are probably extragalactic sources increases at $G\approx 17$~mag. We missed a small number of \textit{Gaia} EDR3 counterparts that lack parallaxes due to processing problems; however, for a large magnitude range, this fraction is only about 1~\% and only slightly increases for very bright and faint sources.

Due to the saturation limit, we further expect most of the RASS counterparts to be brighter than 15~mag and we excluded counterparts fainter than $G=19$~mag to provide some margin for sources detected with a long exposure time or during a flare; again, the influence of the exact cutoff value is negligible. About 1.5~\% of the thusly selected sources do not have magnitude measurements in all three bands. Since these measurements are necessary for our subsequent analysis, we excluded these sources from further analysis. 
We henceforth refer to the selected sources as eligible stellar candidates. 
We then selected those candidate counterparts whose proper motion corrected angular separation from the RASS source is less than five times the positional accuracy of the RASS source.

Since \textit{Gaia} EDR3 is incomplete at the bright end, we also crossmatched the RASS sources with the Tycho2 catalog. Again, we restricted the catalog to Tycho2 sources with magnitude measurements in the $B_T$ and $V_T$ bands and parallaxes from the Hipparcos catalog. From the Tycho2 $B_T$ and $V_T$ magnitudes, we estimated the brightness in \textit{Gaia}'s $G$, $BP$, and $RP$ band adopting the conversion provided in \cite{GaiaDocPhot}. We assume \textit{Gaia} EDR3 and Tycho2 matches to be associated with the same source if their angular separation is smaller than 2~arcsec and their extrapolated G-band magnitude differs by less than 2~mag or the separation is smaller than 5~arcsec and the magnitude difference is less than 0.8~mag; however, the exact values do not influence the result significantly.  

\begin{figure}[t]
        \resizebox{\hsize}{!}{\includegraphics{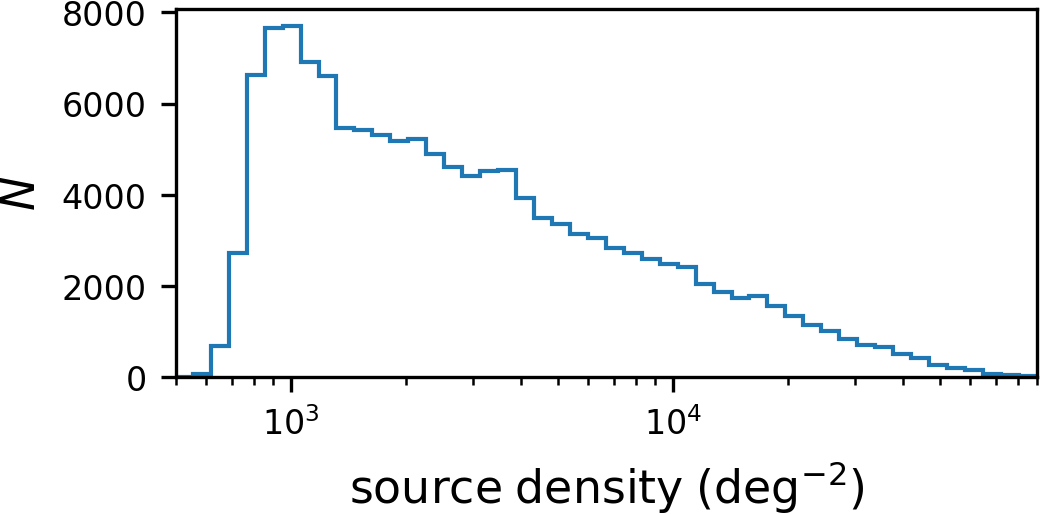}}
        \caption{Distribution of the density of eligible stellar candidates at the RASS positions}
        \label{fig: source density}
\end{figure}

Since the density of the stellar candidates strongly varies between the Galactic plane and poles, the individual counterpart density at the position of the RASS sources also has to be considered in the estimation of the matching probability. Thus, we created an array of Hierarchical Equal Area isoLatitude Pixelation of a sphere (HEALpix, \citet{gorski05}) pixels with a resolution $\sim 27.5$~arcmin and estimated the number of eligible stellar candidates in every pixel. Then, we adopted the number of sources in each pixel, where the RASS source is located, and divided this number by the pixel area to obtain the counterpart density at the location of every RASS source.
For RASS sources located in HEALpix pixels with less than 40 plausible candidate counterparts, we increased the pixel size to obtain a source density that is less affected by statistical fluctuations. The resulting distribution of the source densities at the RASS positions is shown in Fig.~\ref{fig: source density}. The density of eligible stellar candidates is about $10^3~\mathrm{deg^{-2}}$ for most RASS sources, but it increases to a few times $10^4~\mathrm{deg^{-2}}$ for sources that are located close to the Galactic center.

\subsection{Determining the matching probabilities and the stellar fraction}
\label{sec: matching probability}
We deal with the problem of finding the correct identification for $N_X$ X-ray sources, namely the RASS sources, in a catalog of $N_O$ optical counterparts, namely our eligible stellar candidate \textit{Gaia} EDR3 sources. Both the X-ray and optical catalog contain the celestial positions and the positional uncertainties of the sources and they provide additional properties, for example the count rate and the detection likelihood in the X-ray catalog and the magnitude and the parallax in the optical catalog. Due to the submilliarcsecond accuracy of the Gaia positions and proper motions, the uncertainty of the counterpart positions is only a few milliarcseconds after propagating to the RASS observation time. 
Therefore, we assume the errors of the optical positions to be negligible, and we further assume the X-ray positional uncertainties $\sigma_i, i=1, N_X$ to be Rayleigh-distributed. 

The estimation of the matching probabilities in a Bayesian framework is discussed by \citet{schneider21}. The approach compares the hypotheses $H_{ij}, j=1, N_O$ that the i$^\mathrm{th}$ X-ray source is associated with the j$^\mathrm{th}$ counterparts and $H_{i0}$ that the i$^\mathrm{th}$ X-ray source is not associated with any of the counterparts. The prior probabilities that the j$^\mathrm{th}$ optical counterpart is the identification of the i$^\mathrm{th}$ X-ray source and that none of the optical counterparts are associated with the X-ray source is given by 
\begin{eqnarray}
        P(H_{ij}) &=& \frac{p_s p_r}{N_O} = \frac{p_s p_r}{\eta\Omega} \\
        P(H_{i0}) &=& 1 - p_s p_r,
        \label{equ: Prior probabilities}
\end{eqnarray}
where $\eta$ and $\Omega$ are the counterpart density in the vicinity of the X-ray source and the area of the full sky, respectively. Since up to 30~\% of the RASS sources are spurious detections that do not have an optical counterpart, we divided the catalog fraction used by \citet{schneider21} into a stellar fraction $p_s$ and a fraction $p_r$ of real (not spurious) X-ray detections, which is given in Table~1 of \citet{RASS-catalog}. 

Next, data $D_i$, such as the angular separation between the X-ray source and the optical counterpart, were measured, and we estimated the likelihood of obtaining the data $D_i$ considering only the geometric properties and given the hypotheses through 
\begin{eqnarray}
        P(D_i|H_{ij}) &=& \frac{1}{2\pi\sigma_i^2}e^{-\frac{r_{ij}^2}{2\sigma_i^2}} \label{equ: Likelihood data match} \\
        P(D_i|H_{i0}) &=& \frac{1}{4\pi}, 
        \label{equ: Likelihood data no match}
\end{eqnarray}
where $r_{ij}$ is the angular separation between the X-ray source and the optical counterpart. In practice, counterparts with separations much larger than the positional uncertainty can be neglected. 

The fraction of stellar sources in the X-ray catalog $p_s$ is generally not known, but it can be estimated from the likelihood of the matching configuration. Specifically, the likelihood for a single X-ray source is derived by the summation of the hypotheses, and the likelihood of the matching configuration for the full catalog is estimated by the product of the single sources through 
\begin{eqnarray}
        \mathcal{L}_\mathrm{config} &=& \prod_{i=1}^{N_X} \sum_{j=0}^{N_O} P(H_{ij})\times P(D_i|H_{ij}) \\
        &=& \prod_{i=1}^{N_X}\left[ (1 - p_s p_r)\times P(H_{i0}) + \frac{p_s p_r}{\eta\Omega}\times \sum_{j=1}^{N_O} P(D_i|H_{ij})\right].
        \label{equ: Likelihood matching configuration}
\end{eqnarray}
The best value of $p_s$ is estimated by the maximum likelihood through 
\begin{equation}
        \frac{\partial\mathcal{L}_\mathrm{config}}{\partial p_s} = \sum_{i=1}^{N_X} \frac{\frac{p_r}{\eta\Omega}\times \sum_{j=1}^{N_O} P(D_i|H_{ij}) - p_r\times P(H_{i0})}{\sum_{j=0}^{N_O} P(H_{ij})\times P(D_i|H_{ij})} \overset{!}{=} 0.
        \label{equ: Maximum likelihood}
\end{equation}

We applied the stellar fraction $p_s$ resulting from Equation~\ref{equ: Maximum likelihood} to the prior probabilities and obtain the posterior probability that the j$^\mathrm{th}$ counterpart is the correct identification through 
\begin{equation}
p_{ij} = P(H_{ij}|D_i) = \frac{P(D_i|H_{ij})\cdot P(H_{ij})}{\sum_{k=0}^{N_O}P(D_i|H_{ik})\cdot P(H_{ik})}, 
\label{equ: posterior matching probability}
\end{equation}
and the probability that any of the counterparts is the correct identification, and hence the X-ray source is stellar through 
\begin{equation}
p_\mathrm{stellar} = \frac{\sum_{k=1}^{N_O}P(D_i|H_{ik})\cdot P(H_{ik})}{\sum_{k=0}^{N_O}P(D_i|H_{ik})\cdot P(H_{ik})}. 
\label{equ: posterior stellar probability}
\end{equation}

\subsection{Considering additional properties}
\label{sec: considering additional properties}
The estimation described in Sect.~\ref{sec: matching probability} only uses the geometric properties of the counterparts, that is the angular separation, the positional uncertainty, and the counterpart density. However, additional properties can be considered with a Bayes factor $B_{ij}$ by expanding Equation~\ref{equ: Likelihood data match} to 
\begin{equation}
P(D_i|H_{ij}) = \frac{1}{2\pi\sigma_i^2}e^{-\frac{r_{ij}^2}{2\sigma_i^2}}\times B_{ij}.
\label{equ: likelihood any identification Bayes}
\end{equation}
The construction of a Bayes factor, estimated by the fraction of the probability density functions (PDFs) of the considered property for real stellar X-ray identifications and random associations, is described by \citet{schneider21}.

To obtain a clean and unbiased sample of stellar X-ray sources, we selected 846 RASS sources with a geometric matching probability $p_{ij}>0.9$, while 58 of these identifications are expected to be spurious. We inspected these sources individually and filtered out unlikely coronal X-ray emitters due to three nonexclusive categories: 38 and 99 sources are unlikely stellar identifications because of their high X-ray luminosity ($L_X>10^{32}$~erg~s$^{-1}$) and X-ray to G-band flux\footnote{We estimated the G-band flux through $F_G = 10^{-0.4\times G}\times W_\mathrm{eff}\times ZP_\lambda$, adopting the effective bandwidth $W_\mathrm{eff}$ and the zero point $ZP_\lambda$ from \url{http://svo2.cab.inta-csic.es/theory/fps/index.php?mode=browse&gname=GAIA&gname2=GAIA3&asttype=}} ratio (applying an empirical relation\footnote{$\log(F_X/F_G) > \begin{cases} (BP-RP)\times 1.7 - 3.58 & :\; BP-RP<0.7~\mathrm{mag}\\ (BP-RP)\times 0.727 - 2.9 & :\; BP-RP>0.7~\mathrm{mag}\\ \end{cases}$}), respectively, and 67 sources in the sample are located more than 1.5~mag below the main sequence. These counterparts are probably the correct identification of the RASS source, but their X-ray emission is likely produced by a white dwarf or an accreting object and not by a corona. Excluding these unlikely coronal identifications leaves us with 736 sources which we refer to as training set in the following. We expect that a large fraction of the spurious identifications do not pass our filter criteria. We released the catalog of the counterparts with a high geometric matching probability at Centre de Données astronomiques de Strasbourg (CDS) and provide a detailed description in Appendix~\ref{sec: catalog release}.

To be able to compare the properties of the true stellar identifications in the training set with spuriously identified background sources, we shifted all RASS sources randomly between 10 and 20~arcmin and by a random angle and selected all eligible stellar candidates within $5\sigma_r$ to the shifted RASS sources as a control set. 

\begin{figure}[t]
        \includegraphics[width=\hsize]{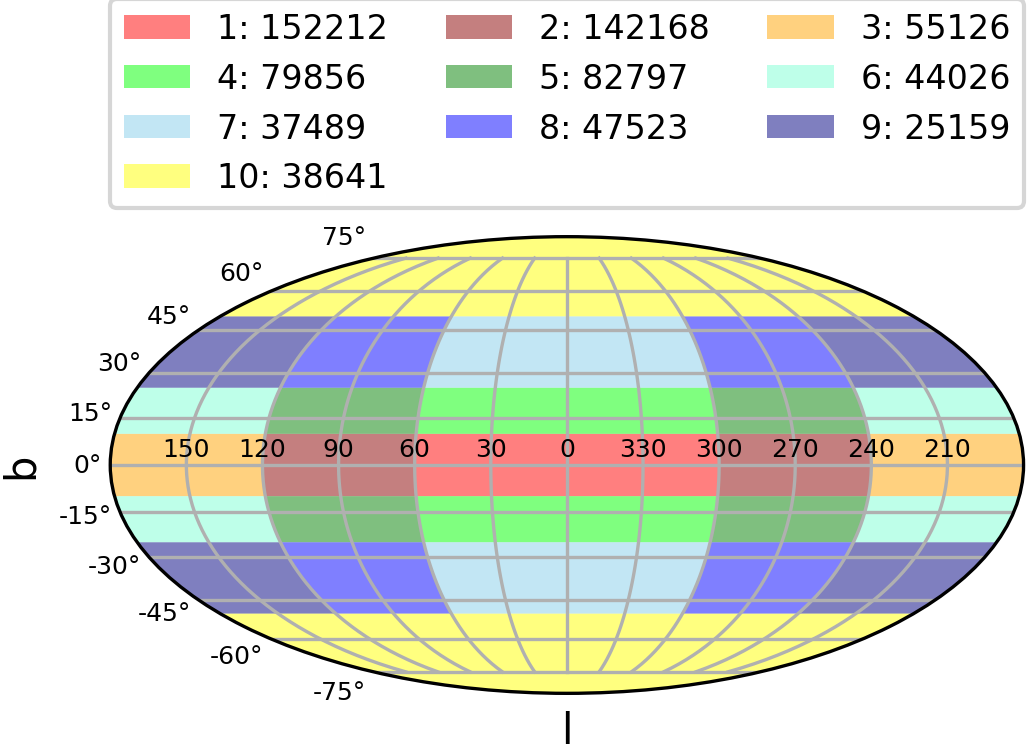}
        \par\bigskip
        \includegraphics[width=\hsize]{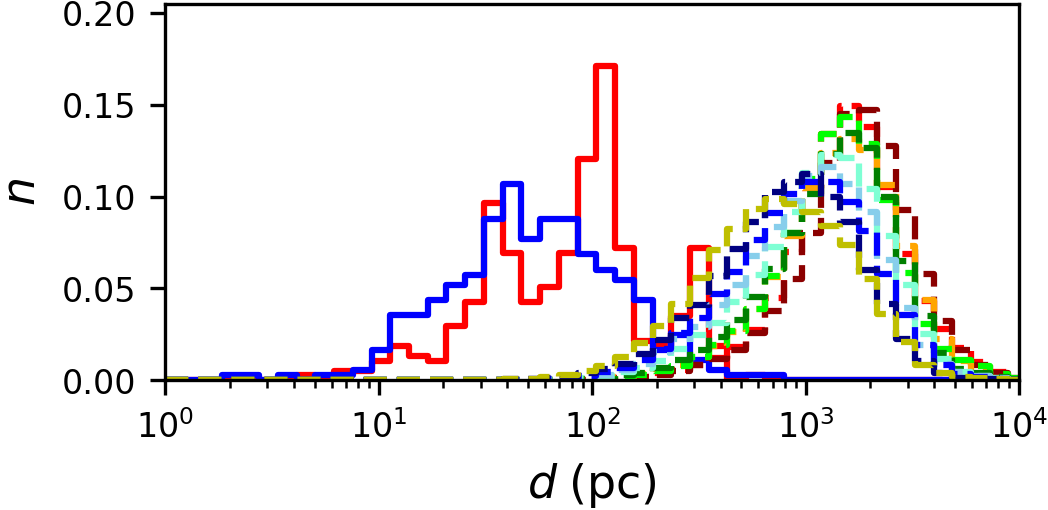}
        \caption{Dependence of the counterpart distances on the Galactic coordinates Top: Bins in which the control set is divided in Galactic coordinates. The legend specifies the number of control set sources in the bins.
        Bottom: Distance distribution of the RASS counterparts at different Galactic coordinates. The solid red and blue histograms show the training set sources with $|b|<25^\circ$ and $|b|>25^\circ$, respectively, and the dashed lines show the distances of the control set sources; the colors refer to the bins shown in the top panel. }
        \label{fig: galactic lat dependence distance}
\end{figure}

Although, there is theoretically no limit to the number of parameters that can be considered in the Bayesian framework, in practice, the training set has to be large enough to cover the n-dimensional parameter space. Since the number of training set sources is rather small, we created two-dimensional Bayes maps of the counterpart distance and the X-ray to G-band flux ratio because the true identifications in the training set and the random associations in the control set substantially differ in these properties.
In Fig.~\ref{fig: galactic lat dependence distance} we show the distance distributions of the training and control set sources at different Galactic coordinates. 
The true identifications of the training set have small distances (typically $<400$~pc), and hence their distance distribution is not affected by the Galactic structure much. On the other hand, the background associations in the control set have much larger distances which increase to lower Galactic latitudes and toward the Galactic center. 
Therefore, we constructed different Bayes maps for each of the bins shown in Fig.~\ref{fig: galactic lat dependence distance}. We applied the control set sources located in the individual bin, but we adopted the whole training set for all bins because the sample size is small and the properties of the training set sources do not change significantly by Galactic coordinates. In App.~\ref{sec: Bayes maps} we show the Bayes maps at different Galactic coordinates. For example sources at 50~pc with $\log(F_X/F_G) = -4$ are upweighted, while counterparts at several kiloparsecs and $\log(F_X/F_G) = -1$ are downweighted. The dividing line between up- and downweighted regions slightly shifts to smaller distances and flux ratios with increasing Galactic latitude, but the Bayes maps overall do not substantially change with Galactic coordinates.

\section{Results}
\label{sec: results}

\subsection{Stellar RASS identifications}
\label{sec: stellar RASS identifications}
We applied our matching procedure to the main RASS catalog and released the resulting catalog electronically at Centre de Données astronomiques de Strasbourg (CDS). In Appendix~\ref{sec: catalog release} we provide a detailed description of the released data.

\begin{figure}[t]
        \includegraphics[width=\hsize]{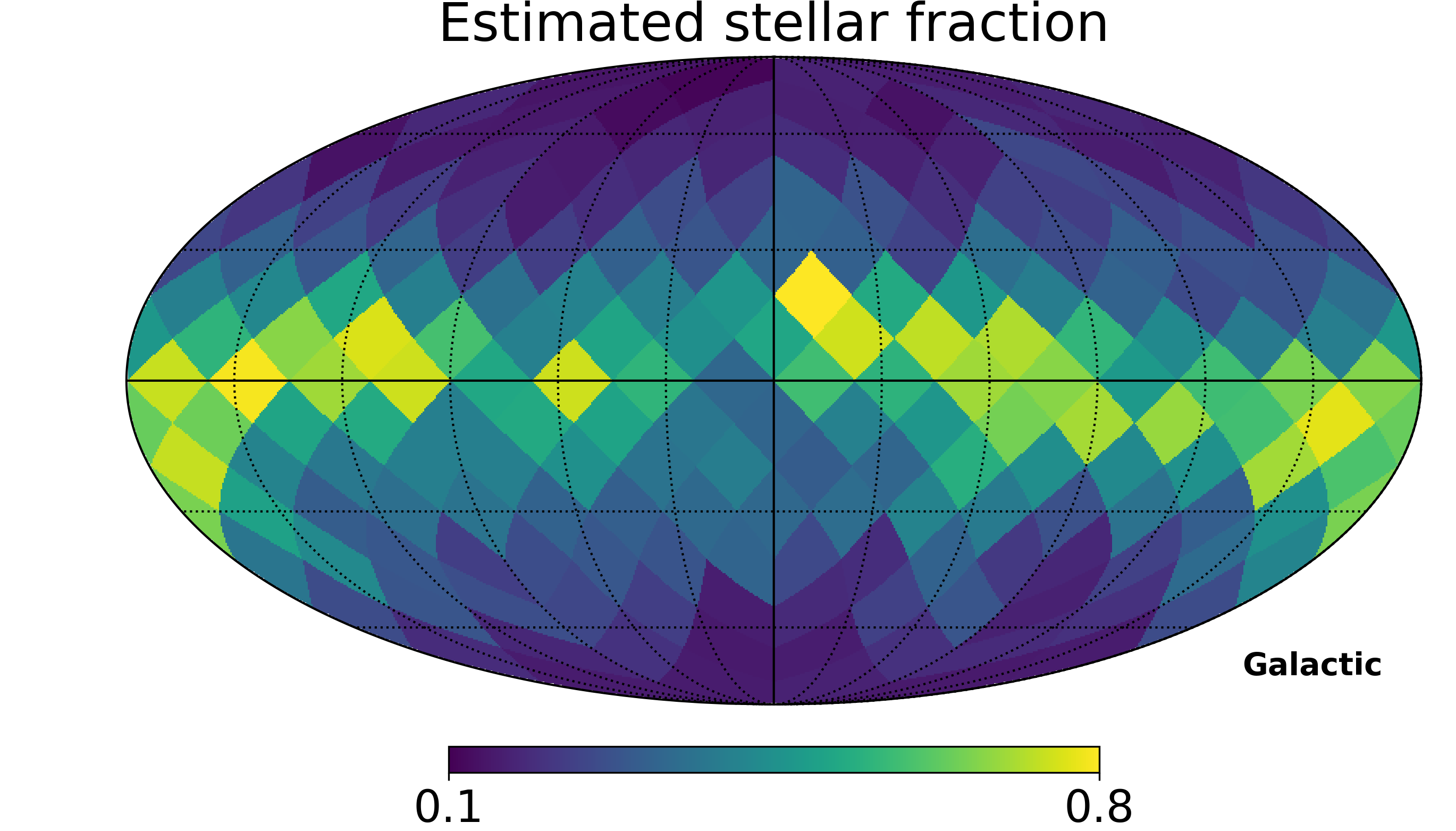}
        \caption{Distribution of the estimated stellar fraction in Galactic coordinates}
        \label{fig: stellar fraction distribution}
\end{figure}

We find 28\,630 (24.9~\%) RASS X-ray sources to be of stellar origin. However, the stellar fraction is not uniformly distributed over the sky; it increases toward the Galactic plane as shown in Fig.~\ref{fig: stellar fraction distribution}. The regions with the largest stellar fractions reach values of up to 0.8 and correspond well with the location of dark nebulae that absorb the X-ray emission of extragalactic objects. The maximal stellar fraction agrees well with the findings of \citet{motch97} who identified,  at a slightly shallower detection limit of $\sim 2.5\cdot10^{-13}\,$erg\,s$^{-1}$\,cm$^{-2}$, 85~\% of their low latitude sample in Cygnus as coronal emitters.

\begin{figure}[t]
        \includegraphics[width=\hsize]{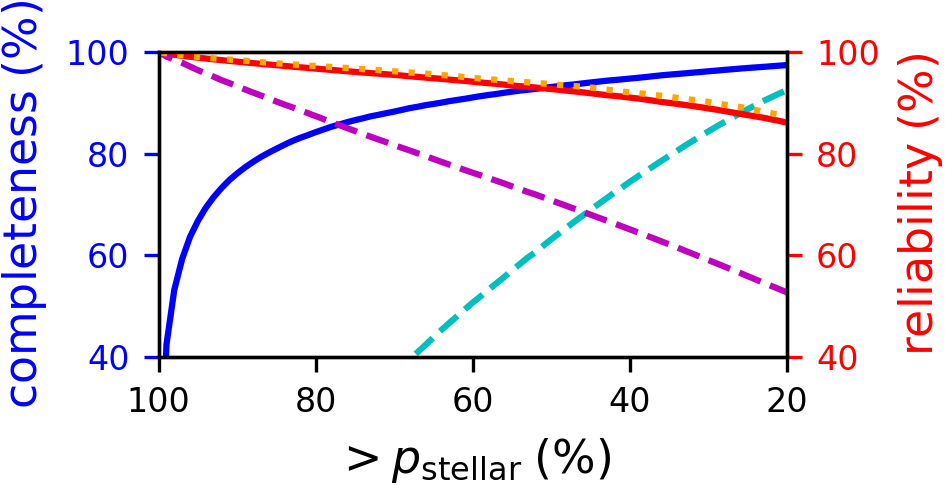}
        \caption{Completeness and reliability of samples with different stellar probability cutoffs. For the dashed cyan and magenta line, only the geometric information is used, while the solid blue and red curves show the completeness and reliability reached when the Bayes map is applied. The dotted orange line shows the reliability expected from shifted RASS sources }
        \label{fig: completeness and reliability}
\end{figure}

With our matching procedure, we can specifically compute the probability $p_{\mathrm{stellar}, i}$ that the i$^\mathrm{th}$ X-ray source is of stellar origin.
With this probability, we can directly estimate the number of missed and spurious identifications through the expressions
\begin{eqnarray}
N_\mathrm{missed} &=& \sum_i^{N_<} p_{\mathrm{stellar}, i} \\
N_\mathrm{spurious} &=& \sum_i^{N_>} (1-p_{\mathrm{stellar}, i}),
\label{equ: number missed spurious}
\end{eqnarray}
as well as the completeness and reliability through the expressions
\begin{eqnarray}
\mathrm{completeness} &=& \frac{N_> - N_\mathrm{spurious}}{N_> - N_\mathrm{spurious} + N_\mathrm{missed}} \\
\mathrm{reliabilty} &=& \frac{N_> - N_\mathrm{spurious}}{N_>},
\label{equ: completeness reliability}
\end{eqnarray}
where $N_>$ and $N_<$ are the number of sources above and below a stellar probability cutoff. 
In Fig.~\ref{fig: completeness and reliability} we show the resulting completeness and reliability of our stellar RASS identifications for samples with different stellar probability cutoffs. Naturally, a high cutoff value leads to a very reliable but incomplete sample, while relaxing the cutoff criterion increases the completeness at the cost of a lower reliability. At $p_\mathrm{stellar} = 0.51$, the correct number of stellar sources is recovered and the completeness and reliability intersect at about 92.9~\%. This is a significant improvement over the consideration of the geometric properties only, which reach a completeness and reliability of only about 68~\%.. The number of spurious associations can be tested independently by applying the matching procedure to randomly shifted RASS sources' counterparts which are spurious by definition. The thusly obtained estimate for the reliability fits the values derived from the stellar probability very well (see dotted orange line in Fig.~\ref{fig: completeness and reliability}), confirming the high accuracy of our estimated reliability.

Not all of the stellar RASS identifications can be unambiguously associated with one \textit{Gaia} EDR3 stellar candidate; instead, 3621 RASS sources have two, 81 have three, and four have four plausible stellar counterparts with a matching probability of $p_{ij} > 0.1$. Most of these sources are multiple star systems, and the component responsible for the X-ray emission cannot be identified because of the large positional uncertainties of the RASS sources. In fact, the RASS source is likely  a superposition of the emission from multiple components. 

A comparison with previous identifications by \citet{salvato18} using NWAY shows that about one-third of our stellar identifications are located in sky areas not considered by NWAY, which is focused on the identifications of extragalactic X-ray sources. Among the sources that the catalog used by \citet{salvato18} and our main catalog have in common, about 92~\% of the RASS sources are identified with the same counterpart by both methods. On the other hand, NWAY provides stellar counterparts to 15.6~\% of the RASS sources that are not associated with stars by our procedure. A closer inspection of these sources reveals that about 29~\% of them are not part of our stellar candidate list either because the parallax is missing 
        in Gaia EDR3 or the stated counterpart is obviously an extragalactic object despite its stellar classification of \citet{salvato18}; a detailed comparison is provided in Appendix~\ref{sec: comparison NWAY}.

\subsection{Comparison with the stellar eRASS1 sources}
\label{sec: comparison stellar eRASS1}

\begin{figure}[t]
        \includegraphics[width=\hsize]{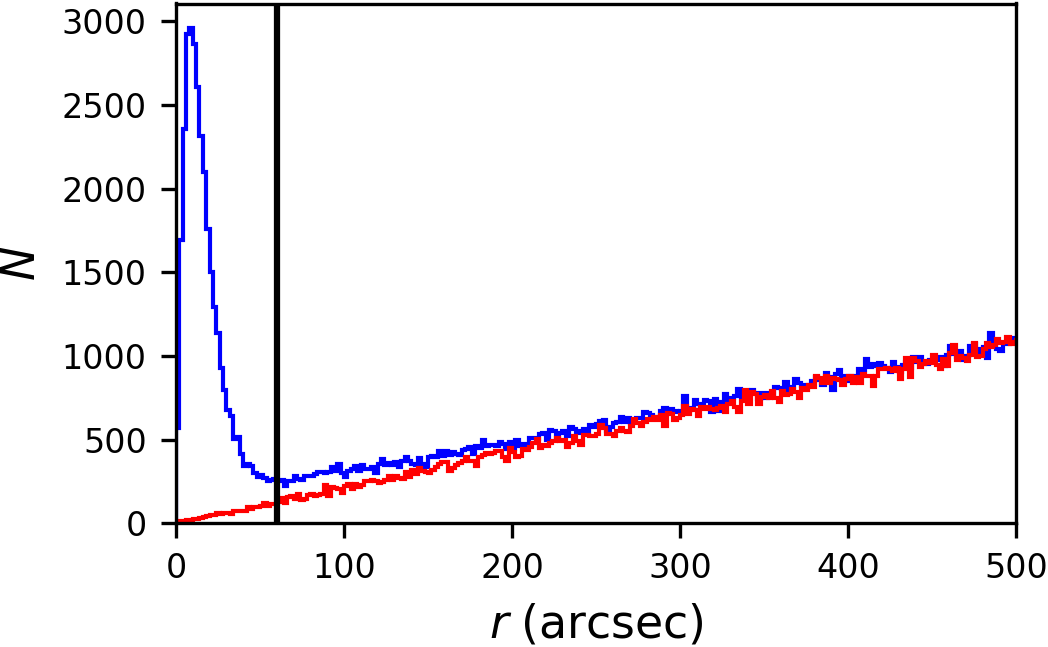}
        \caption{Distribution of the angular separations between the RASS sources and the eRASS1 counterparts for the real (blue) and shifted (red) RASS sources. The vertical line shows our cutoff at 60~arcsec}
        \label{fig: RASS eRASS separations}
\end{figure}

To investigate the reliability of our stellar RASS identifications, we consider X-ray sources detected during eRASS1. The effective point response function of eROSITA
        and, hence, the positional accuracy provided by eRASS1 is typically a factor of 8 better than that provided by RASS for sources detected in both surveys (cf. Sect.~\ref{sec: RASS catalog} and \ref{sec: eRASS catalog}). Therefore, the eRASS1 sources are ideally suited to carry out reliability tests of our stellar RASS identifications.  
We specifically crossmatched the 52\,842 RASS sources located at $l>180^\circ$ with the eRASS1 catalog and show the distribution of the angular separations in Fig.~\ref{fig: RASS eRASS separations}. At small separations, the Gaussian peak of the true identifications is clearly visible, while the distribution approaches the linear slope of random associations at larger separations. As a compromise between spurious and missed identifications, we selected 32\,363 RASS-eRASS1 associations within an angular distance of 60~arcsec. We expect that the RASS sources not detected in eRASS1 are either spurious sources or detected during a flare. Indeed, some of the RASS sources without an eRASS1 identification show larger excess variances than those sources detected in both surveys. We identify 12\,757 of the RASS sources with an eRASS1 counterpart as stellar; 12\,692 sources have counterparts near or above the main sequence and, thus, they are likely coronal X-ray emitters (see Sect.~\ref{sec: CMD}). 

To validate the stellar identifications of the RASS-eRASS1 associations,
we adopted preliminary stellar identifications of the eRASS1 catalog (Freund et al. (in prep.)). These identifications  are obtained by a two-step Bayesian algorithm similar to the method described in \citet{schneider21} and the procedure applied in this paper. First, the eRASS1 sources are positionally crossmatched with eligible Gaia EDR3 counterparts considering the angular separation, positional uncertainty, counterpart density, and catalog fraction. In the second step, the counterparts are weighted according to their properties compared to the properties of a training and control set (cf. Sect.~\ref{sec: matching probability} and \ref{sec: considering additional properties}). In contrast to the method presented here, for the preliminary
stellar identifications of eRASS1, we adopted a Bayes map as a function of the X-ray to G-band flux ratio and the $BP-RP$ color because these properties vary less as a function of Galactic latitude. For the comparatively bright RASS-eRASS1 associations, we expect a completeness and reliability of about 97.5~\% of the stellar eRASS identifications.

\begin{figure}[t]
        \includegraphics[width=\hsize]{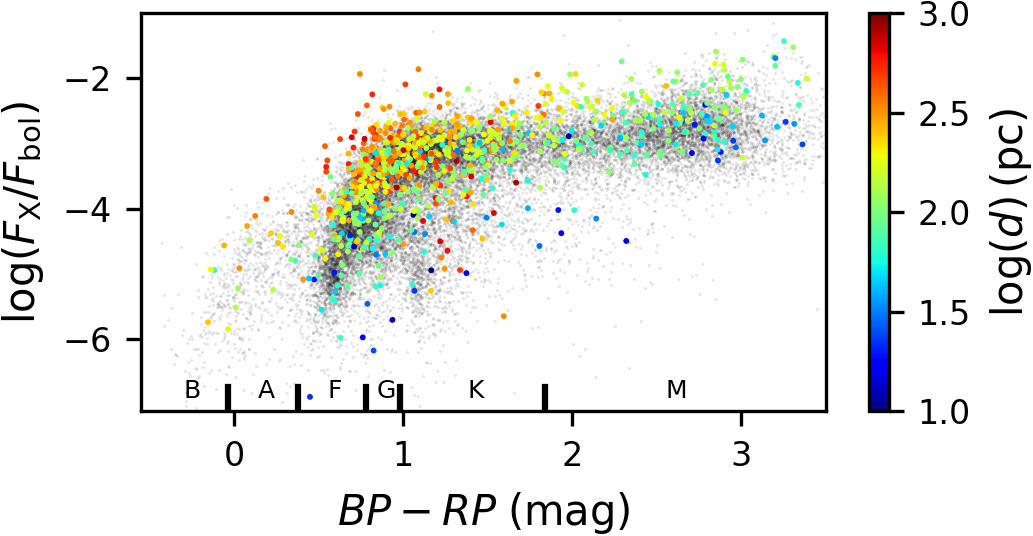}
        \caption{$F_X/F_\mathrm{bol}$ vs. $BP-RP$ color for the sources identified as stellar by RASS and not by eRASS1. The color scales  with the distance of the counterparts and the gray dots show the distributions for the whole stellar RASS sample as a comparison. At the bottom we show the ranges of the spectral types as guidance.}
        \label{fig: RASS not confirmed eRASS F_X/F_bol}
\end{figure}  
\begin{figure}[t]
        \includegraphics[width=\hsize]{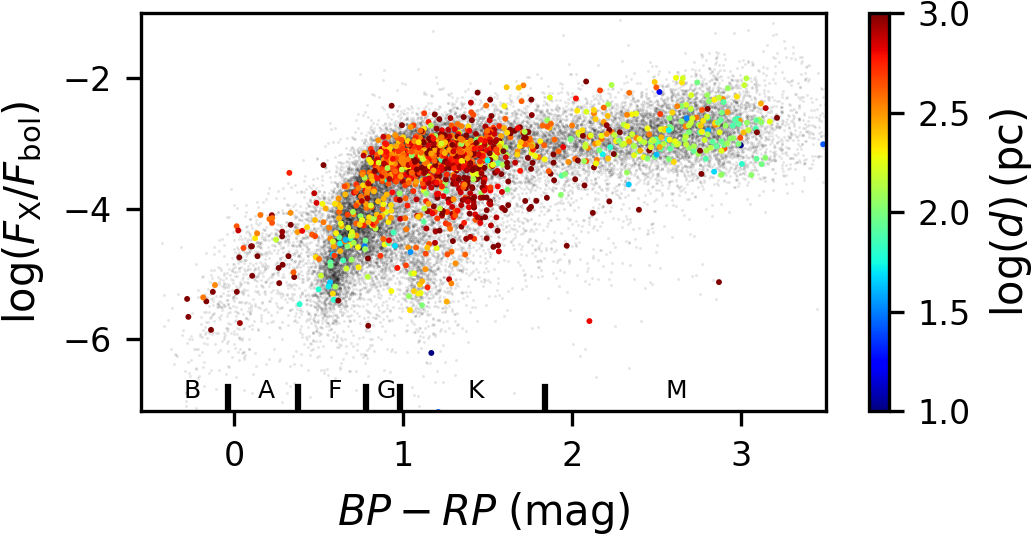}
        \caption{$F_X/F_\mathrm{bol}$ vs. $BP-RP$ color for the sources identified as stellar by eRASS1 and not by RASS. The color scales  with the distance of the counterparts and the gray dots show the distributions for the whole stellar RASS sample as a comparison.}
        \label{fig: RASS missed eRASS F_X/F_bol}
\end{figure}  
The stellar identification based on RASS has been confirmed by eROSITA for 11\,811 (93~\%) of the sources, which is the predicted reliability of our stellar RASS sample. In Fig.~\ref{fig: RASS not confirmed eRASS F_X/F_bol} we show the X-ray to bolometric flux\footnote{For the bolometric correction and the conversion 
between the $BP-RP$ and other photometric colors, we adopt throughout this paper the values given in a table based on \citet{pec13} which are available at \url{http://www.pas.rochester.edu/~emamajek/EEM_dwarf_UBVIJHK_colors_Teff.txt} (Version 2021.03.02)} ratio as a function of the $BP-RP$ color for the stellar RASS identifications not confirmed by eRASS1 (cf Sect.~\ref{sec: F_X/F_bol}). The spurious identifications not confirmed by eRASS1 tend to have slightly larger $F_X/F_\mathrm{bol}$ values, especially a few F- and G-type sources above the saturation limit are not confirmed. These RASS sources have been identified because color is not considered by the weighting scheme. However, the sources not confirmed have overall similar properties as the correctly identified sources. On the other hand, 1291 (10.2~\%) sources are classified as stellar by eRASS1, but not by RASS, and we show the RASS X-ray bolometric flux ratios of these sources in Fig.~\ref{fig: RASS missed eRASS F_X/F_bol}; subsequently, the completeness is probably slightly lower than estimated from the stellar probabilities. This is caused by the fact that some RASS sources have larger positional offsets than expected from the given uncertainties (see Sect.~\ref{sec: Positional uncertainties of the RASS sources}), and furthermore some of the eRASS1 identifications might be spurious. 
For example, some of these counterparts have large distances ($>1$~kpc) which might be erroneously associated with the X-ray source because the preliminary identification procedure for eRASS1 currently does not consider the counterpart distances.

\section{Astrophysical properties of the stellar RASS sources}
\label{sec: properties of stellar RASS sources}
In the following we discuss the properties of the stellar RASS sources. To avoid ambiguous identifications, we restricted the analysis to 28\,097 stellar RASS sources with $p_\mathrm{stellar} > 0.51$ and $p_{ij} > 0.5$.

\subsection{Color-magnitude diagram}
\label{sec: CMD}
\begin{figure*}[t]
        \includegraphics[width=17cm]{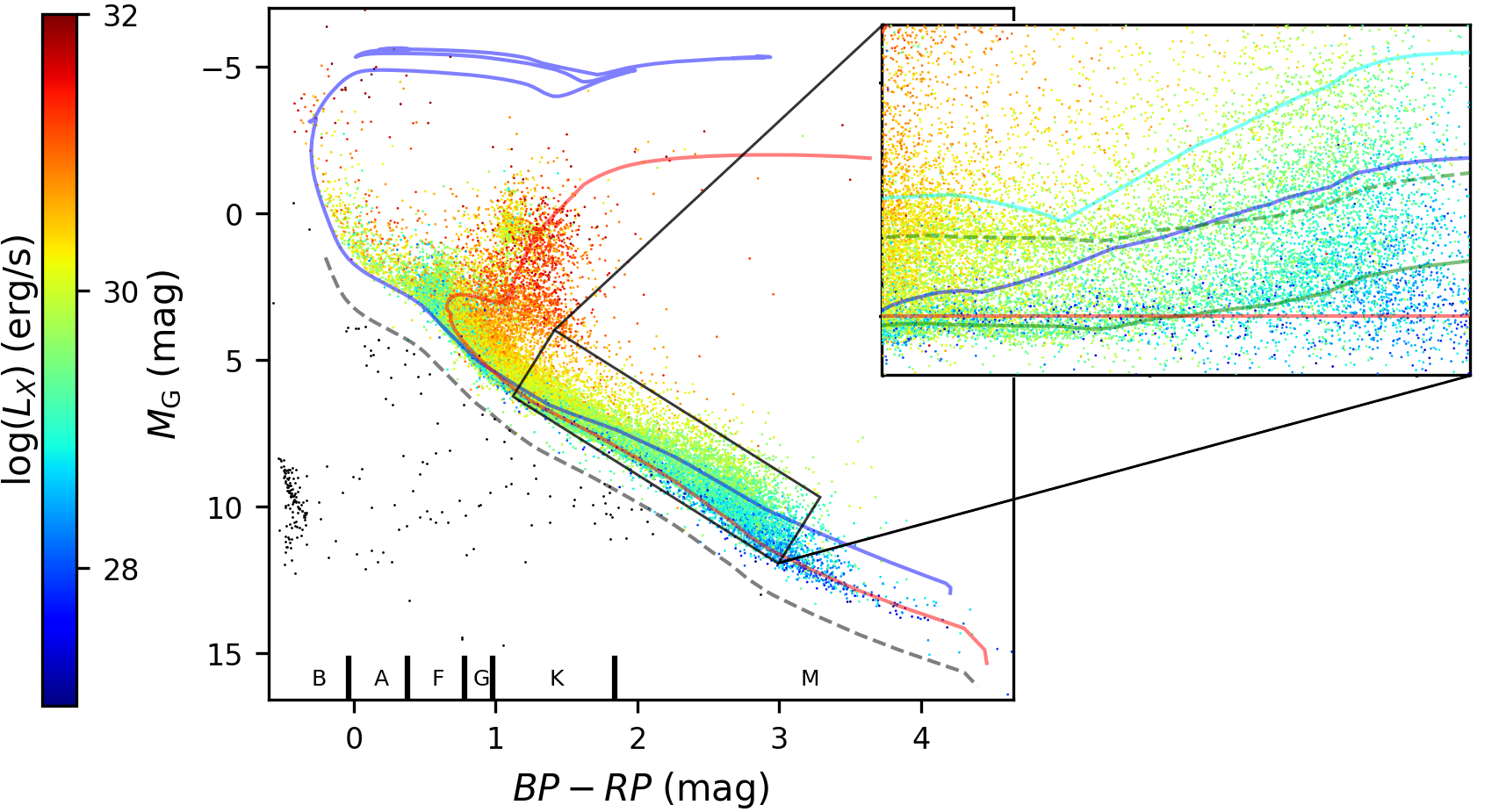}
        \caption{Color-magnitude diagram of the RASS sources identified as stars. The color scales with the X-ray luminosities, and the blue and red solid lines show the PARSEC isochrones for stellar ages of $4\times 10^7$ and $4\times 10^9$ years \citep{bressan12}, respectively. The cyan and green solid lines in the close-up indicate isochrones of $1.25\times 10^7$ and $1.25\times 10^8$ years and the green dashed line shows the binary sequence at $1.25\times 10^8$ years. Sources located below the dashed line (black dots) are more than 1.5~mag fainter than the main sequence.}
        \label{fig: CMD}
\end{figure*}

In Fig.~\ref{fig: CMD} we show the color-magnitude diagram (CMD) for the RASS sources identified as stars. Our X-ray selected sample contains dwarfs from late M-type to B-type. The number of counterparts with $BP-RP > 3$~mag (corresponding to a spectral type of M4V) is very small. These sources are very faint, and hence unlikely to be detected in RASS. Furthermore, their number in our training set is also small (18 out of 736), and therefore very red dwarfs might be slightly underrepresented due to our Bayes map. We also find some O-type stars in the sample, but they have rather large $BP-RP$ colors due to reddening, and hence they are located in the regime of B-, A-, and F-type stars with absolute magnitudes typically larger than \hbox{-3}~mag. We note that no attempt to correct for extinction has been made in either the optical or in the X-ray because most stellar RASS sources are located within a few hundred parsecs. However, premain sequence and O-type stars located in molecular clouds and sources at large distances, for example giants and early-type stars, can be substantially affected by reddening. 

Many of the late-type stellar RASS sources are located at positions in the CMD that are compatible with young stars with ages of a few $10^7$~yr. This is a result of the X-ray luminosity decreasing with age \citep{skumanich72,preibisch05}. Some of the stellar RASS sources might also be binaries that are unresolved in \textit{Gaia} EDR3 and shifted to lower magnitudes in Fig.~\ref{fig: CMD}. The sources above the main sequence have steadily increasing X-ray luminosities, which is caused by a higher activity and by the fact that the higher bolometric luminosity allows for higher X-ray luminosity due to the saturation limit. 

Also the giant branch is clearly visible in Fig.~\ref{fig: CMD}. Many of the X-ray detected giants have large luminosities of more than $1\times 10^{31}$~erg~s$^{-1}$, and we expect these sources to be active binaries such as RS CVn systems. We find some RASS counterparts in the red giant branch beyond the so-called dividing line and we inspected some of these sources individually. A few sources are known spectroscopic binaries, hence the X-ray emission might be produced by a late-type companion. However, in particular, many of the objects with large X-ray luminosities are known symbiotic binaries where the X-ray emission is produced by an accretion process \citep[cf.][]{huensch98c}.

Compared to the 3407 RASS-Hipparcos (RasHip) sample presented by \citet{guillout99}, our CMD is much better populated. We especially find more giants, early-type sources, and young stars because the RasHip sample is biased to sources brighter than magnitude 9 and closer than 100~pc.

We also find 223 sources that are located more than 1.5~mag below the main sequence and we have excluded these sources from the following analysis. We expect that most of these counterparts are the correct identification of the RASS source, but that the X-ray emission is unlikely to be produced by a coronal source. 
Instead, the X-ray emission is either very soft and produced by a white dwarf or cataclysmic variables are likely responsible when the X-ray emission is rather hard. Although a detailed study of these sources is beyond the scope of this paper, they might be interesting in another context. Therefore, we included these sources with a flag in our released catalog (see Appendix~\ref{sec: catalog release}). However, a complete analysis of the compact objects detected in RASS requires an adapted Bayes map.

\subsection{X-ray to bolometric flux ratio}
\label{sec: F_X/F_bol}
\begin{figure*}[t]
        \centering
        \includegraphics[width=12cm]{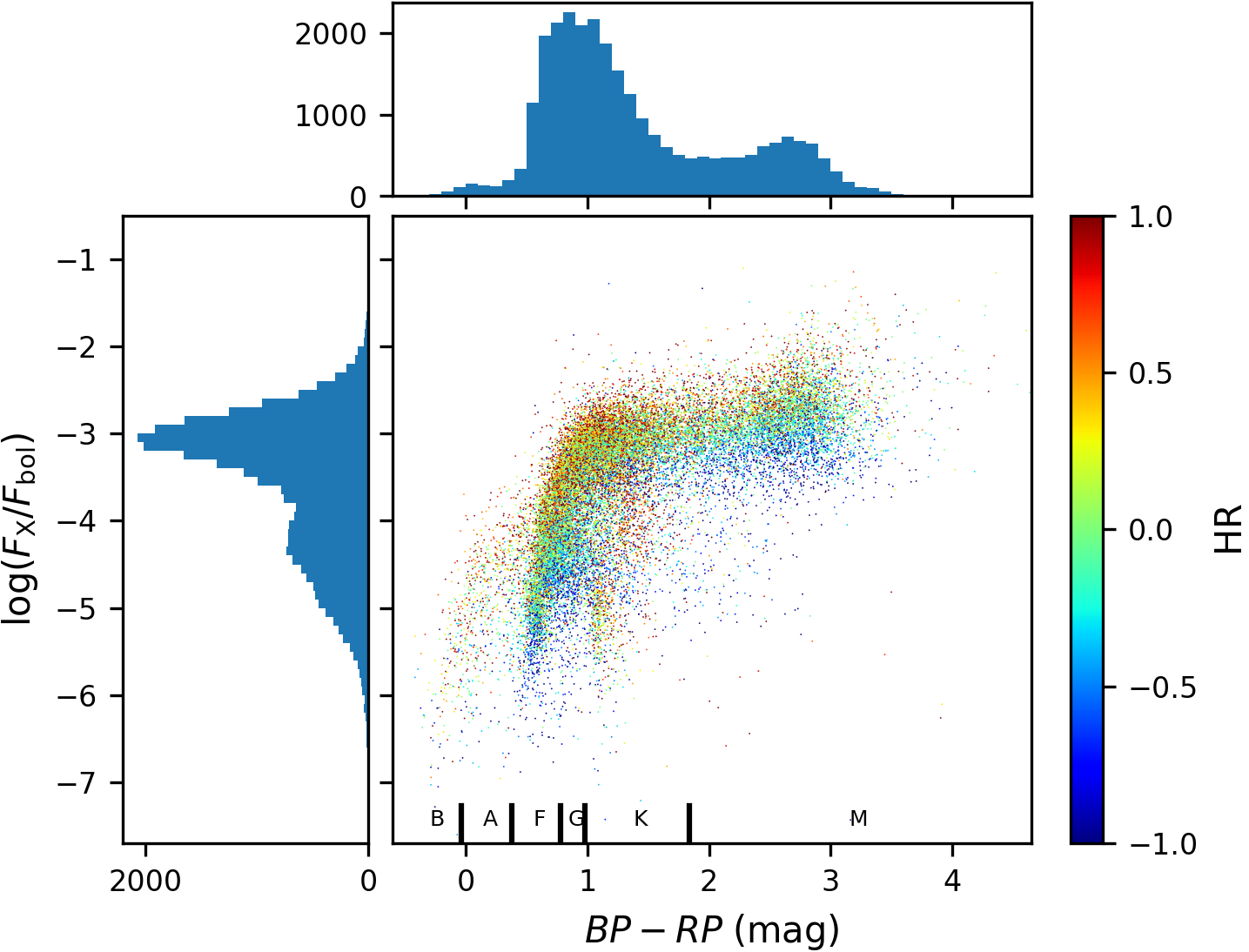}
        \caption{$F_X/F_\mathrm{bol}$ vs. $BP-RP$ color for the RASS sources identified as stars. The color scales with the hardness ratio of the RASS sources. The histograms show the distribution of the X-ray to bolometric flux ratios and the $BP-RP$ color, respectively.}
\label{fig: F_X/F_bol}
\end{figure*}

We show the fraction of the X-ray to bolometric fluxes as a function of the $BP-RP$ color for the stellar RASS sources in Fig.~\ref{fig: F_X/F_bol}. At spectral types of early F, where stars begin to develop an outer convection zone, the number of sources and the $F_X/F_\mathrm{bol}$ rapidly increases. We also find some A-type counterparts, although especially early A-type stars are generally not expected to produce X-ray emission \citep{schmitt85}, here, a late-type companion is likely responsible for the X-ray emission in most cases. 

While some stellar RASS sources have X-ray to bolometric flux ratios below $10^{-7}$, the $F_X/F_\mathrm{bol}$ distribution peaks around the saturation limit at $10^{-3}$ and the distribution rapidly decreases for larger flux ratios. The width of the decrease is defined by the intrinsic scatter of the saturation limit and by flares during the RASS detection. Very few sources are found with $F_X/F_\mathrm{bol} > 10^{-2}$; they are likely detected during a large flare, and furthermore the fraction of spurious identifications is larger for these sources than for the whole sample. However, the fraction of very active sources is smaller than in the sample of \citet{freund18} because of the longer exposure times of RASS compared to XMMSL. The hardness ratio changes with the X-ray activity; sources with a high flux ratio exhibit rather hard X-ray emission and low active sources emit soft X-rays.

\begin{figure}[t]
        \includegraphics[width=\hsize]{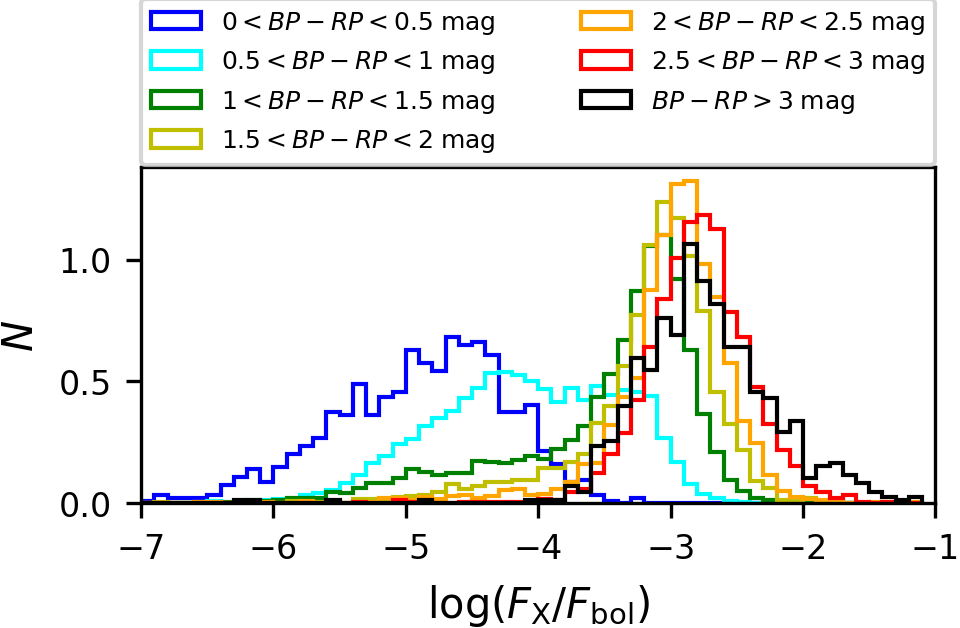}
        \caption{Comparison of the $F_X/F_\mathrm{bol}$ distribution for the stellar RASS sources with different $BP-RP$ colors}
        \label{fig: color comparison F_X/F_bol}
\end{figure}

In Fig.~\ref{fig: color comparison F_X/F_bol} we compare the distributions of the X-ray to bolometric flux ratios for different colors. The bluest sources have very low flux ratios because they have no or very shallow convection zones and their X-ray emission is often produced by a late-type companion. Also very few of the F- and G-type stars reach flux ratios of $10^{-3}$ because only the youngest stars are saturated. Some of the sources with $1<BP-RP<1.5$~mag have low flux ratios, most of them are giants, but very few later-type stars at small $F_X/F_\mathrm{bol}$ values are detected in RASS due to the detection limit. Instead, the distribution of the stellar RASS sources later than $BP-RP=1$~mag show a strong peak around $F_X/F_\mathrm{bol} \approx 10^{-3}$. The peak of the distributions and the highest flux ratios are shifted to higher values for later-type stars. The reason is most likely that the late-type sources flare more frequently and the flares have a larger impact on the $F_X/F_\mathrm{bol}$ values due to their small X-ray and bolometric fluxes in quiescence. We note that the mean fractional X-ray fluxes derived by \citet{agueros09} are smaller
by up to 0.6~dex, especially for K-type sources. However, their sample is smaller by a factor of about 60  and contains fewer optically faint sources than our sample.

\subsection{Color distribution of detected and not-detected stellar sources}
\label{sec: color comparison}

\begin{figure}[t]
        \includegraphics[width=\hsize]{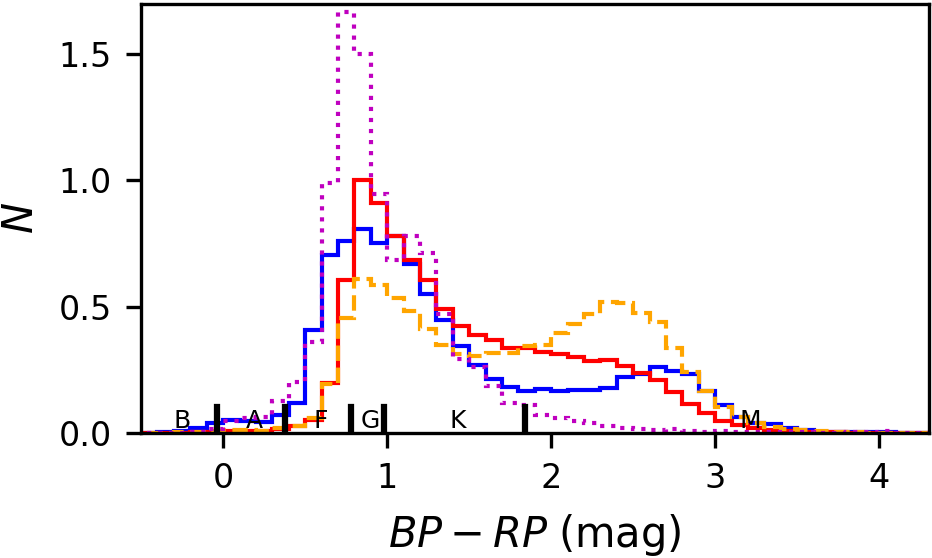}
        \caption{Comparison between the normed $BP-RP$ color distributions of the stellar candidates identified with a RASS source (blue) and the stellar candidates not associated with a RASS source (red). The dashed orange and dotted magenta histograms show the color distributions for the unassociated candidates within 800~pc and those brighter than $G=13$~mag, respectively. }
        \label{fig: color comparison X-ray stars background}
\end{figure}

To compare the properties of the stellar RASS sources with stellar sources not associated with a RASS detection, we selected stellar candidates of the control set  located close to a randomly shifted RASS sources (see Sect.~\ref{sec: considering additional properties}). The counterparts of the shifted sources represent stellar sources that do not emit X-rays at a level detectable in RASS. In Fig.~\ref{fig: color comparison X-ray stars background} we compare the $BP-RP$ color distributions of the stellar RASS sources with the counterparts to shifted RASS sources. The sources detected in RASS have properties significantly different from that of the undetected stellar candidates. First, both distributions show a sharp increase for F-type stars but while this increase is located at early F-type for the detected sources, the number of uncorrelated sources increases at late F-type. Furthermore, the true stellar RASS sources show a broad second peak around spectral type M3 that is not visible for the counterparts to the shifted sources. 

The stellar candidates identified with a RASS source are generally brighter and have smaller distances than the undetected sources. Therefore, we show in Fig.~\ref{fig: color comparison X-ray stars background} the $BP-RP$ color distributions of the uncorrelated sources brighter than $G=13$~mag and closer than 800~pc. The bright sources show a sharp increase at early F-type as the sources detected in RASS and the close stellar candidates have a bimodal distribution as the stellar RASS sources, although the distribution of the red stars peaks at slightly earlier spectral types. We note that a combination of bright and nearby stellar candidates does not fit the color distribution of the stellar RASS sources, and therefore we conclude that the properties of the stellar candidates detected in RASS significantly differ from the sources with lower X-ray emission that are not detected in RASS. 

\subsection{Three-dimensional distribution}
\begin{figure*}[t]
        \centering
        \includegraphics[width=1\textwidth]{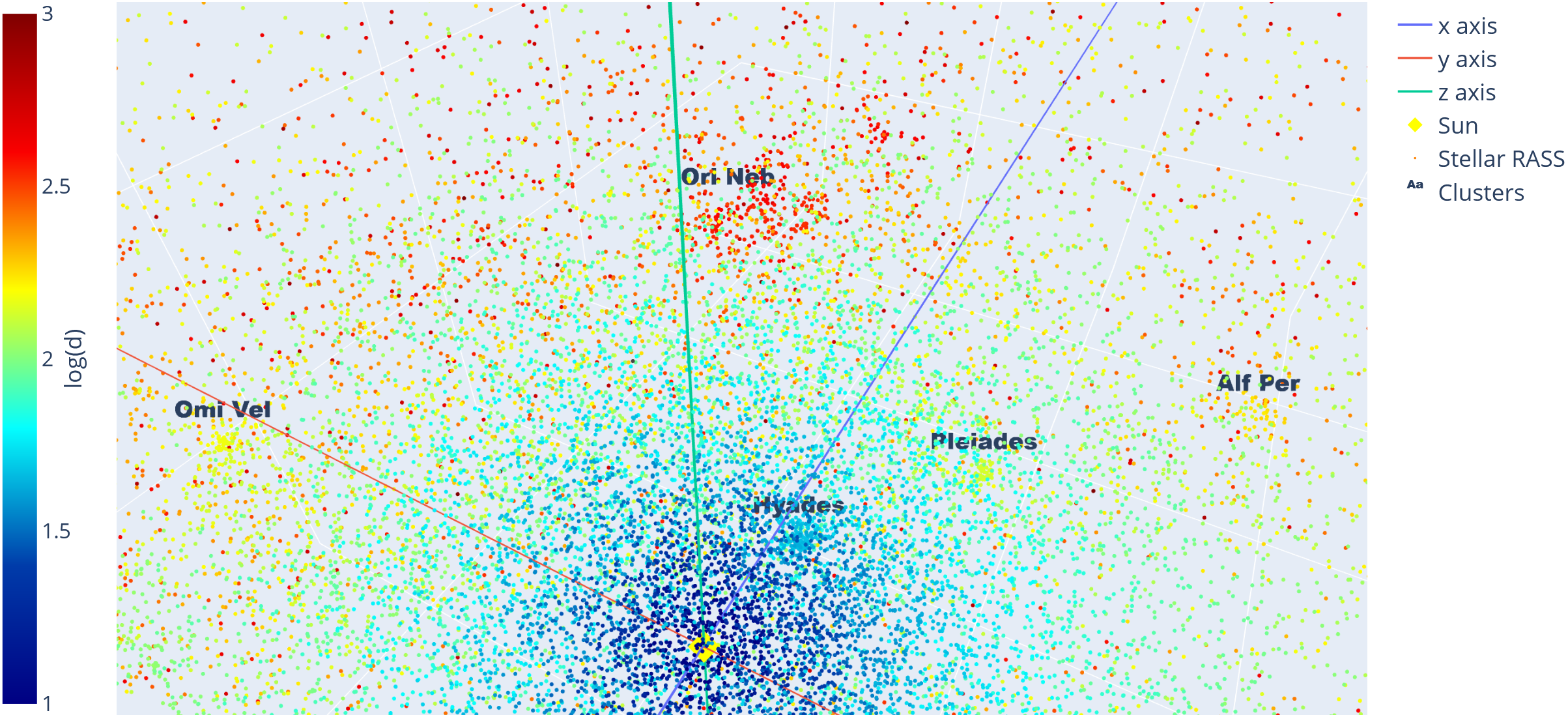}
        \caption{Close-up of the three-dimensional distribution of the stellar RASS sources in Galactic Cartesian coordinates. From the point of view of the plot, we are looking from the Galactic north pole on the Galactic plane (green z axis), along the blue x axis in the direction of the Galactic anticenter and the red y axis is directed against the Galactic rotation to the upper left. The color scales with the distances of the stellar RASS sources to the Sun, whose position is shown by the yellow diamond.}
        \label{fig: 3D distribution close-up}
\end{figure*}
We estimated Galactic Cartesian coordinates for the stellar RASS sources from their \textit{Gaia} positions and parallaxes; an interactive representation of the three-dimensional distribution of the stellar RASS sources is available online\footnote{Link will be added (TBD)}. 
It shows that most of the stellar RASS sources are located within 600~pc to the Sun; however, there are some outliers with distances of more than 1~kpc, but it is important to note that most of them have quite low stellar probabilities (typically $p_\mathrm{stellar}<0.7$), and hence the fraction of spurious identifications is rather large. Due to their low distances from the Sun, the distribution of the stellar RASS sources is almost spherical with only a slight flattening caused by the thickness of the Galactic plane. The source density strongly increases with decreasing distance to the Sun due to the RASS detection limit.

When zooming in, several open clusters are visible. In Fig.~\ref{fig: 3D distribution close-up} we show a two-dimensional projection of a region closer to the Sun; here, the Hyades, Pleiades, $\alpha$ Persei, and $o$ Velorum clusters are visible and the density of stellar RASS sources is clearly enhanced in these regions. The Orion Nebula is located in the background at the top of Fig.~\ref{fig: 3D distribution close-up}; in that region, the RASS sources are clustered in many small structures.

\citet{guillout98a,guillout98b} studied the spatial distribution of a sample of RASS
X-ray sources identified with Tycho counterparts with regard to the Gould
        Belt, an ellipsoidal structure with semi-major and semi-minor axes of about 500
        and 340 pc, which surrounds the Sun and appears to contain many star forming
        regions and young stellar objects. The authors found the distribution
        to be compatible with a Gould-disk, a structure extending inward from the Gould Belt,
        but being disrupted around the Sun. However, in contrast to the sample presented here, \citet{guillout98a,guillout98b} only had parallaxes for their brightest  counterparts.
        
\begin{figure*}[t]
        \centering
        \includegraphics[width=14cm]{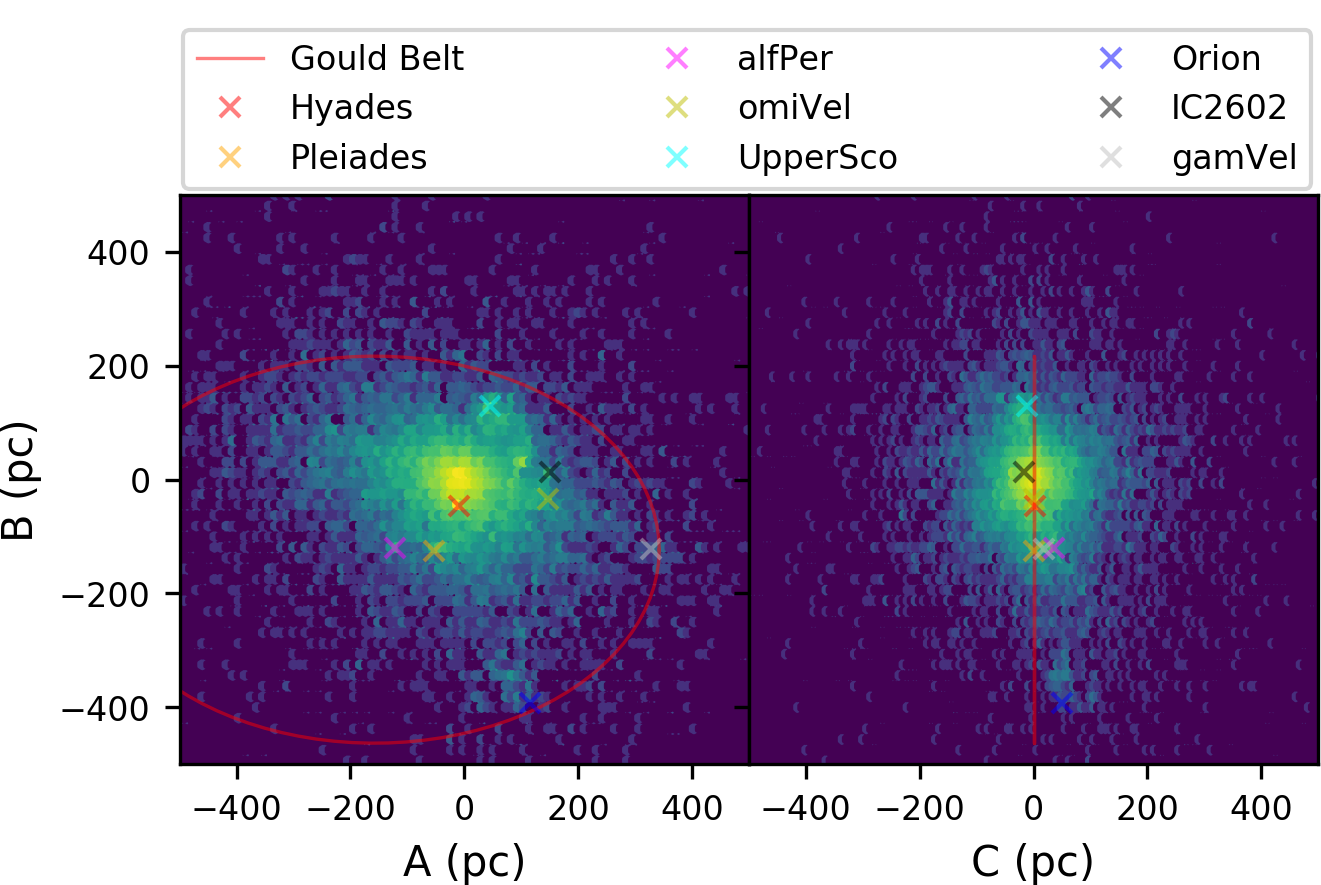}
        \caption{Density of the stellar RASS sources with respect to the Gould Belt. The A and B axes are located in the plane of the Gould Belt and directed as the semi-major and semi-minor axis, the C axis is directed perpendicular to the Gould Belt. The color scales with the density of the stellar RASS sources. The red line shows the position of the Gould Belt as defined by \citet{guillout98b} and the crosses indicate the positions of known stellar clusters. }
        \label{fig: Gould belt coordinates}
\end{figure*}

Adopting the geometry described by \citet{guillout98b}, we show
        in Fig.~\ref{fig: Gould belt coordinates} the surface density of our Gaia-identified stellar RASS sources derived from the three-dimensional spatial information
        projected, first, on the plane defined by the Gould Belt (left panel) and, second, an edge-on density view (right panel).
        The stellar RASS sources are not uniformly
        distributed around the Sun. Instead, many known stellar
        clusters are visible, and the distribution in the plane of the Gould Belt (Fig.~\ref{fig: Gould belt coordinates}, left)
        is elongated in a direction consistent with the alleged shape of the Gould Disk.
        The farther parts of the structure, corresponding to the lower left part in the left panel of Fig.~\ref{fig: Gould belt coordinates},
        are beyond the sensitivity limits of RASS. The source density in the edge-on view is
        also compatible with a concentration of sources in the plane of the Gould Belt.
        Therefore, we find the three-dimensional spatial distribution of the stellar
        RASS sources geometrically compatible with the
        Gould-disk structure described by \citet{guillout98b}.  


\section{Conclusion}
\label{sec: conclusion}
In this paper we have presented the first identification attempt of the whole stellar content detected in the \textit{ROSAT} all-sky survey. It is an all-sky X-ray flux limited sample that is, to our knowledge, the largest sample of stellar X-ray sources presented so far. 
We adopted an automatic procedure to identify the RASS sources with stellar candidate counterparts selected from bright ($G<19$~mag) \textit{Gaia} EDR3 sources with a parallax significance $>3\sigma$. Our procedure considers geometric information of the match (angular separation, positional accuracy of the RASS sources, and counterpart density) as well as additional properties, namely the counterpart distances and the X-ray to G-band flux ratios.

A crossmatch with preliminary eRASS1 sources shows that the positional offsets of some RASS sources are larger than expected for Gaussian distributed positional uncertainties. This deviation increases with the positional uncertainty given in the RASS catalog, and hence we discuss in this paper only the identifications of the 115\,000 RASS sources with the highest positional accuracies. We also publish a supplementary catalog with the stellar counterparts of the remaining RASS sources, but for those sources our identifications are less reliable.

Of the highly accurate RASS sources, we identify 28\,630 (24.9~\%) as stars and provide a stellar probability for each RASS source. From these probabilities, we estimate that our identifications are to about 93~\% complete and reliable. We confirmed this value by comparisons with identifications to randomly shifted RASS sources, preliminary stellar identifications of eRASS1 sources, and the results from \citet{salvato18}. Thus, we might have missed a small number of stellar RASS sources due to incompletenesses in \textit{Gaia} EDR3, a problem we expect to be reduced or remedied by future \textit{Gaia} releases, and furthermore some counterparts to the RASS sources have larger positional offsets than expected from Gaussian distributed uncertainties. Previous identifications from \citet{salvato18} generally agree well with our results, but the stellar classification by \citet{salvato18} that is solely based on X-ray and AllWISE fluxes seems to be less reliable compared to our selection of stellar candidates. In contrast to \citet{salvato18}, we also provide stellar identifications in the crowded regions of the Galactic plane and near the Large and Small Magellanic Clouds where the association is more difficult, and hence we identify about 35~\% more stellar RASS X-ray sources. 

The CMD of the stellar RASS sources shows all basic features known from an optically selected CMD. The main sequence contains dwarfs of spectral types from late M- to B-type; the number of counterparts with spectral types later than M4 are very rare due to their faintness. We also find many sources in the giant branch but for most of the identified red giants, the X-ray emission is probably produced by a late-type companion or a compact accreting object. Many of the stellar RASS sources are young  stars with ages of a few $10^7$ years according to their position in the CMD or their absolute magnitudes are increased by an unresolved companion. The X-ray to bolometric flux ratios strongly increase for early F-type stars due to the onset of convection and near $F_X/F_\mathrm{bol} = 10^{-3}$ the saturation limit is clearly visible. The peak value of the $F_X/F_\mathrm{bol}$ distribution and the maximum values increase to later spectral types because these sources have stronger and more frequent flares. The color distribution of the stellar candidates detected in RASS clearly differ from the undetected sources. Thanks to the parallaxes of \textit{Gaia} EDR3, we can for the first time access the three-dimensional distribution of the stellar RASS sources. Overall the stellar RASS sources are nearly spherically distributed; on smaller scales, stellar clusters with increased source densities are clearly visible.

The ongoing eROSITA all-sky survey is expected to detect more than 20 times as many stellar X-ray sources as identified in RASS. With an identification algorithm similar to the one presented in this paper, we are confident to be able to identify the stellar eRASS sources. The results presented in this paper will provide the definitive reference to eROSITA stellar science and will help to make optimal use of the unprecedented potential of eROSITA.

\begin{acknowledgements} 
        We thank Mara Salvato for useful comments and suggestions.

        SF gratefully acknowledge supports through the Integrationsamt Hildesheim, the ZAV of Bundesagentur f\"ur Arbeit, and the Hamburg University, SC by DFG under grant CZ 222/5-1, JR by DLR under grant 50 QR 2105, and PCS by DLR under grant 50 OR 1901 and 50 OR 2102. SF thanks Gabriele Uth and Maria Theresa Lehmann for their support.

        This work has made use of data from the European Space Agency (ESA) mission
        {\it Gaia} (\url{https://www.cosmos.esa.int/gaia}), processed by the {\it Gaia}
        Data Processing and Analysis Consortium (DPAC,
        has been provided by national institutions, in particular the institutions
        participating in the {\it Gaia} Multilateral Agreement.

        This work is based on data from eROSITA, the soft X-ray instrument  aboard  SRG,  a  joint Russian-German  science  mission  supported  by  the  Russian  Space  Agency (Roskosmos), in the interests of the Russian Academy of Sciences represented by its Space Research Institute (IKI), and the Deutsches Zentrum für Luft- und Raumfahrt  (DLR).  The  SRG  spacecraft  was  built  by  Lavochkin  Association (NPOL)  and  its  subcontractors,  and  is  operated  by  NPOL  with  support  from IKI and the Max Planck Institute for Extraterrestrial Physics (MPE). The development and construction of the eROSITA X-ray instrument was led by MPE, with  contributions  from  the  Dr.  Karl  Remeis  Observatory  Bamberg  \&  ECAP (FAU  Erlangen-N\"urnberg),  the University  of  Hamburg  Observatory,  the Leibniz Institute for Astrophysics Potsdam (AIP), and the Institute for Astronomyand Astrophysics of the University of T\"ubingen, with the support of DLR and the Max Planck Society.  The Argelander Institute for Astronomy of the University of Bonn and the Ludwig Maximilians Universit\"at Munich also participated in the science preparation for eROSITA.

        This research has made use of the SIMBAD database, operated at CDS, Strasbourg, France.
\end{acknowledgements}

\bibliographystyle{aa} 
\bibliography{mybib}

\begin{appendix}
\section{Bayes maps}
\label{sec: Bayes maps}
In Fig.~\ref{fig: Bayes maps} we show the Bayes maps at the different Galactic coordinates. We applied a Gaussian kernel density estimator to estimate the PDFs of the training and control set and adopted a rather large bandwidth compared to Scott's rule to better smooth statistical fluctuations in our training set. Furthermore, we added a small constant to the PDFs so that the Bayes factor approaches unity in regions sparsely populated by training and control set sources. For example, optically very bright sources with a low X-ray to G-band flux ratio at large distances or sources that are very near to the Sun but very faint in the optical are very rare in the training and control set, and hence we did not weigh these sources. Due to sparse populations, also the weighting for very bright and near sources as well as faint and large distant sources is reduced which might seem unphysical. However, since the number of such candidates is extremely small, this has little to no influence on our stellar identifications. The same is true for the other details of the estimation of the Bayes map.
\begin{figure*}[t]
        \subcaptionbox{Bin 1}{\includegraphics[width=8.5cm]{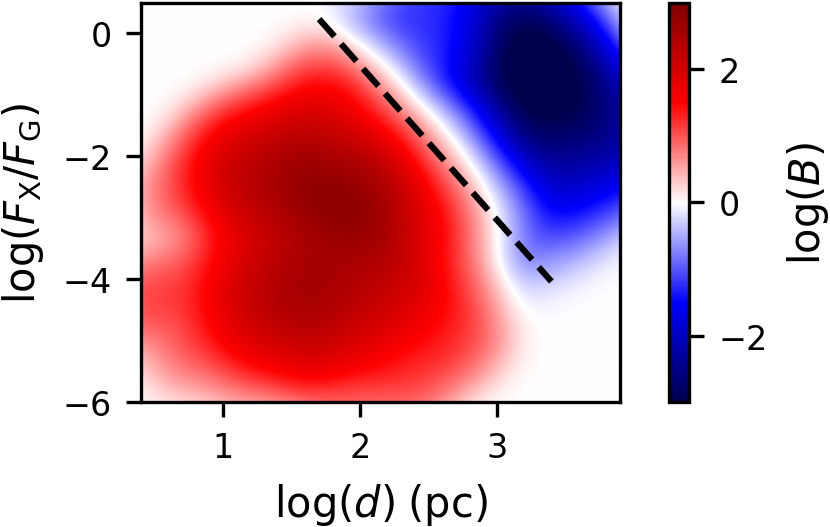}}
        \hfill
        \subcaptionbox{Bin 2}{\includegraphics[width=8.5cm]{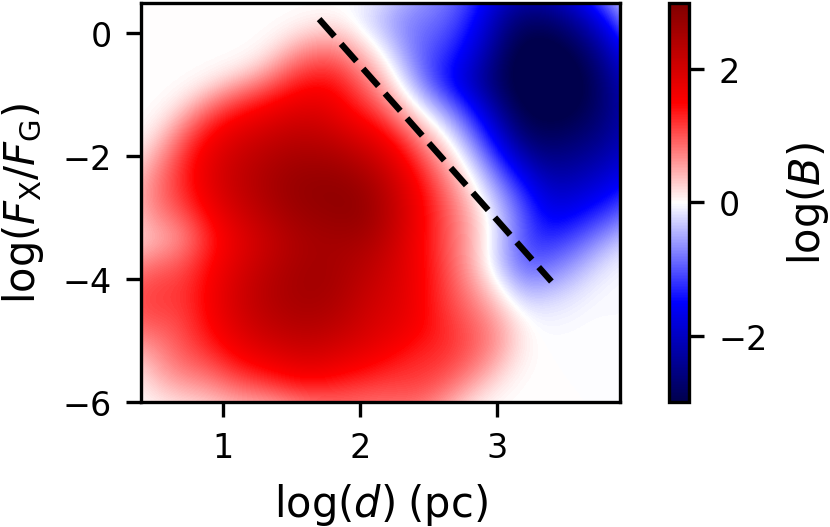}}
        \par\bigskip
        \subcaptionbox{Bin 3}{\includegraphics[width=8.5cm]{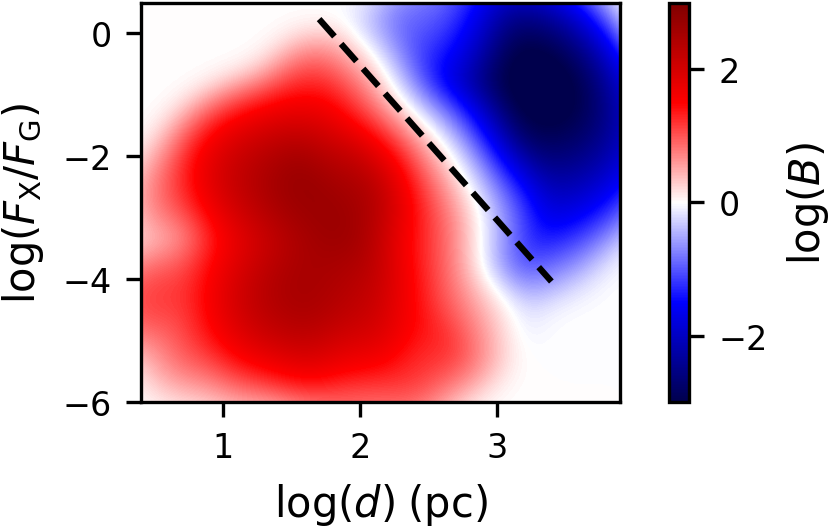}}
        \hfill
        \subcaptionbox{Bin 4}{\includegraphics[width=8.5cm]{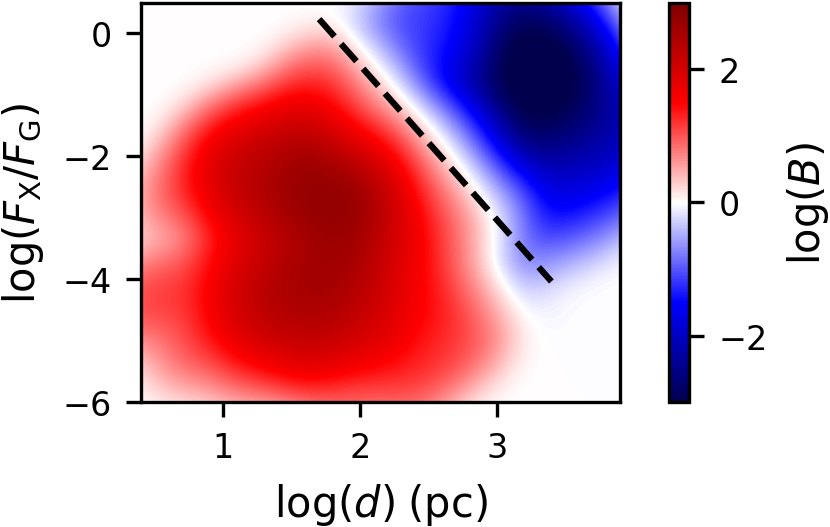}}
        \par\bigskip
        \subcaptionbox{Bin 5}{\includegraphics[width=8.5cm]{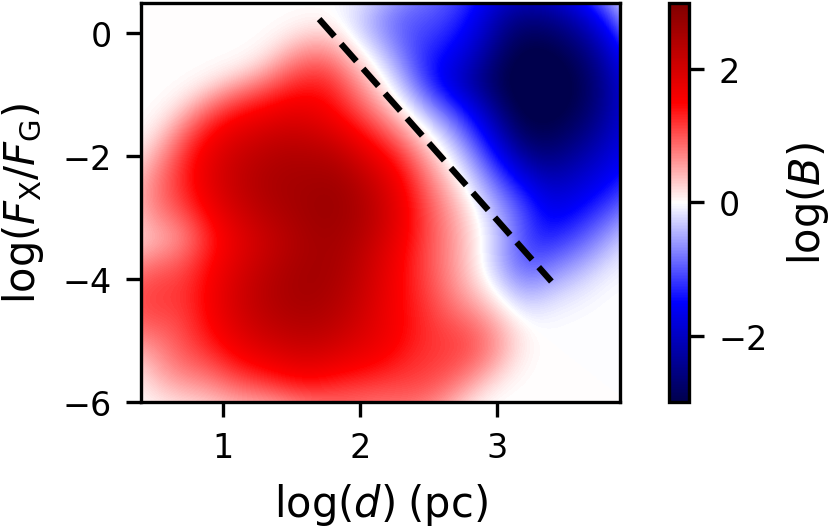}}
        \hfill
        \subcaptionbox{Bin 6}{\includegraphics[width=8.5cm]{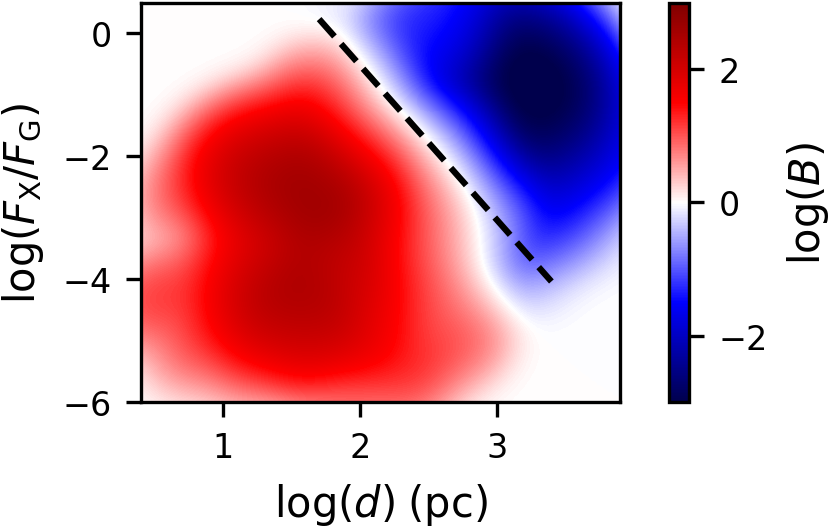}}
        \caption{Bayes maps for the bins shown in Fig.~\ref{fig: galactic lat dependence distance}. The color scales with the Bayes factor. The dashed line is plotted at the same position in every panel as a reference.} 
        \label{fig: Bayes maps}
\end{figure*} 
\begin{figure*}[t]\ContinuedFloat
        \subcaptionbox{Bin 7}{\includegraphics[width=8.5cm]{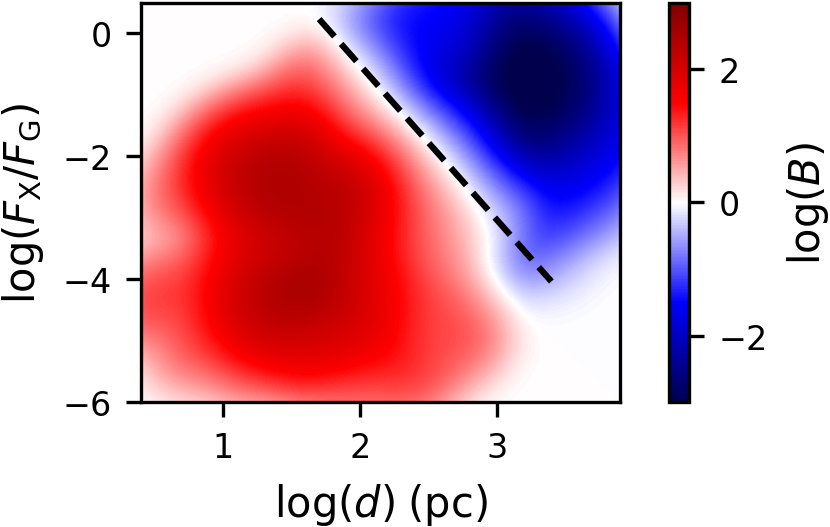}}
        \hfill
        \subcaptionbox{Bin 8}{\includegraphics[width=8.5cm]{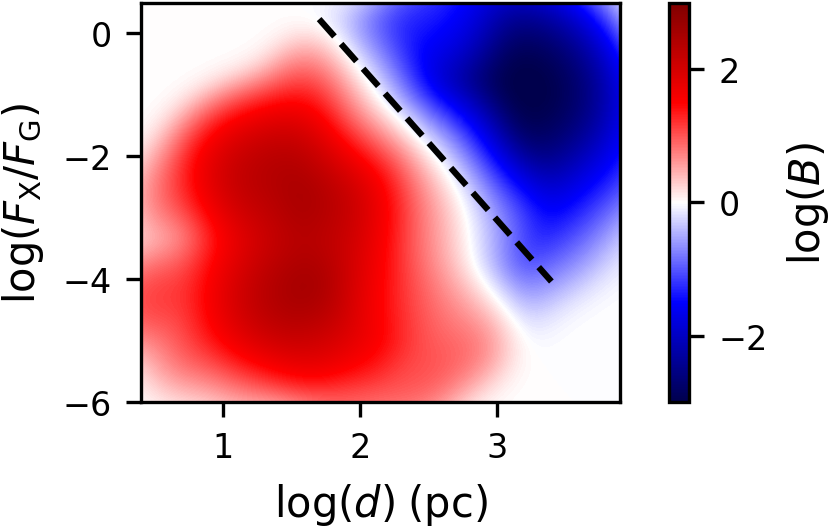}}
        \par\bigskip
        \subcaptionbox{Bin 9}{\includegraphics[width=8.5cm]{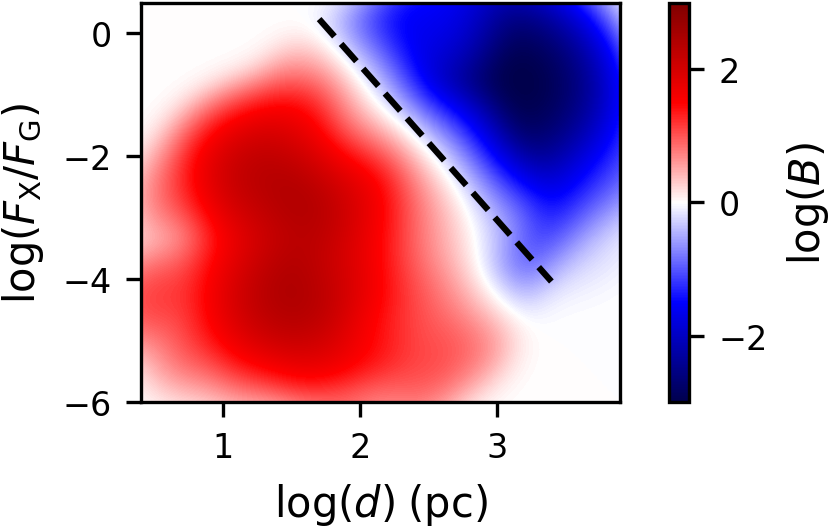}}
        \hfill
        \subcaptionbox{Bin 10}{\includegraphics[width=8.5cm]{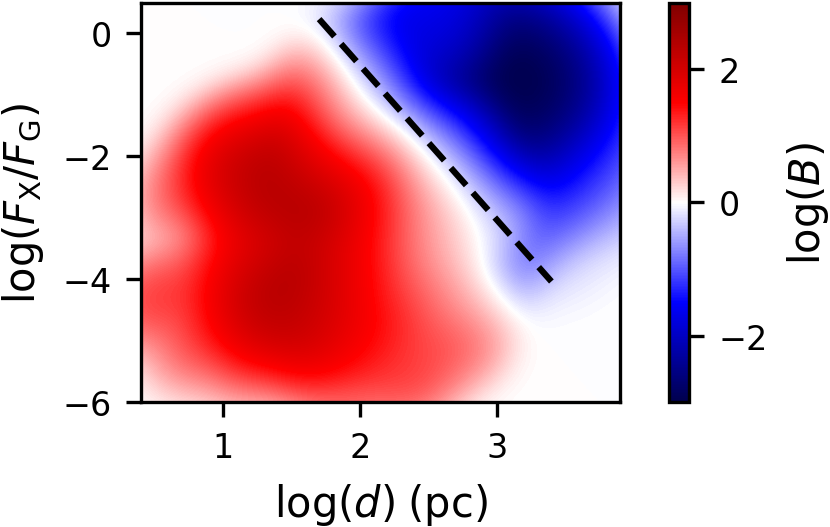}}
        \caption{continued} 
        \label{fig: Bayes maps 2}
\end{figure*} 

\newpage~\newpage~\newpage~\newpage~\newpage~\newpage
\section{Catalog release}
\label{sec: catalog release}
In Table~\ref{tab: matching table} we provide the first 11 entries of our main matching catalog adopted throughout this paper. The full catalog is available electronically at Centre de Données astronomiques de Strasbourg (CDS) containing the stellar counterparts with a matching probability $p_{ij} > 0.1$ to all RASS sources with a high quality positional accuracy and a stellar probability $p_\mathrm{stellar} > 0.2$. A supplementary catalog with the counterpart of the RASS sources with a low positional uncertainty is also available at CDS, but for these sources our identification procedure is less reliable. 

The catalog contains the names of the RASS sources and the stellar matches (\textit{Gaia} source ID or Tycho2 ID if the source is not available in \textit{Gaia} EDR3), the positional uncertainties of the RASS sources estimated by Equation~\ref{equ: positional uncertainty}, and the matching separations between the RASS sources and the stellar identifications. Furthermore, we provide the stellar probabilities (p\_stellar) and the matching probabilities (p\_ij) of the individual counterparts. Table~\ref{tab: matching table} also lists the proper motion corrected coordinates, the RASS X-ray fluxes, the RASS hardness ratio, the G-band magnitudes, the $BP-RP$ colors, and the parallaxes of the counterparts. Sources located more than 1.5~mag below the main sequence are flagged as a subdwarf.

RASS sources with multiple plausible counterparts have multiple entries in the catalog. For the discussion of the properties of the stellar RASS sources, we adopted the counterparts with $p_\mathrm{stellar} > 0.51$ and $p_{ij} > 0.5$ of the main catalog. 
\begin{sidewaystable}
        \caption{Basic properties of the stellar counterparts to the RASS sources}
        \label{tab: matching table}
        \centering
        \resizebox{\textwidth}{!}{
        \begin{tabular}{llllllllllllll}
                \hline \hline
                RASS\_NAME & match\_ID & sigma\_r & sep & p\_stellar & p\_ij & match\_RA & match\_Dec & Fx & HR & G & BP\_RP & plx & subdwarf \\
                &  & arcsec & arcsec &  &  & deg & deg & erg/s/cm2 &  & mag & mag & mas &  \\
                \hline
                2RXS J111220.4+872612 & 1151502176829452672 & 10.34 & 17.11 & 0.9571 & 0.9571 & 168.185944 & 87.435271 & 4.55e-14 & -1.00 & 13.04 & 1.21 & 4.01 & False \\
                2RXS J121506.0+874154 & 1728674141358771840 & 6.02 & 10.24 & 0.9996 & 0.9996 & 183.836210 & 87.699983 & 1.01e-12 & -0.16 & 6.19 & 0.46 & 20.67 & False \\
                2RXS J095440.9+872818 & 1151646212853601536 & 17.40 & 24.46 & 0.8186 & 0.8176 & 148.516947 & 87.472113 & 3.34e-13 & 0.99 & 14.61 & 2.73 & 11.61 & False \\
                2RXS J144219.2+871438 & 1728461317139734016 & 11.00 & 25.33 & 0.5936 & 0.5908 & 220.697854 & 87.239816 & 1.82e-13 & 0.04 & 13.49 & 1.32 & 4.14 & False \\
                2RXS J123746.9+875804 & 1728789727518860416 & 13.05 & 17.83 & 0.9917 & 0.9917 & 189.430636 & 87.972964 & 2.37e-13 & 0.14 & 8.99 & 0.65 & 8.61 & False \\
                2RXS J154734.6+872205 & 1728088578401606400 & 17.53 & 3.78 & 0.9757 & 0.9757 & 236.912721 & 87.367554 & 7.35e-14 & -0.66 & 12.47 & 1.39 & 8.28 & False \\
                2RXS J105207.8+884320 & 1152608349951908992 & 13.55 & 11.37 & 0.9905 & 0.9902 & 162.909618 & 88.724061 & 5.99e-14 & -1.00 & 12.02 & 1.31 & 6.89 & False \\
                2RXS J160911.9+875830 & 1728963450356277632 & 8.15 & 10.48 & 0.9951 & 0.9951 & 242.240334 & 87.977098 & 1.28e-13 & -0.81 & 8.91 & 0.76 & 10.39 & False \\
                2RXS J162322.1+883607 & 1729043474187041536 & 11.79 & 21.85 & 0.7308 & 0.7308 & 245.654162 & 88.606030 & 9.56e-14 & nan & 8.15 & 1.05 & 3.34 & False \\
                2RXS J064236.2+880446 & 1152331242957967104 & 6.84 & 2.53 & 0.9978 & 0.7838 & 100.640436 & 88.078970 & 1.28e-12 & 0.23 & 10.69 & 0.87 & 5.78 & False \\
                2RXS J064236.2+880446 & 1152331238662558464 & 6.84 & 4.62 & 0.9978 & 0.2141 & 100.678522 & 88.078695 & 1.28e-12 & 0.23 & 12.22 & 1.10 & 5.55 & False \\
                \hline
        \end{tabular}
        }
\tablefoot{The full table is available at CDS}
\end{sidewaystable}

In Table~\ref{tab: training set table} we present the first 11 entries of the counterparts with a geometric matching probability $>0.9$. In addition to Table~\ref{tab: matching table}, the geometric stellar (p\_geo) and matching probabilities (p\_ij\_geo) are provided. Sources with a high X-ray to bolometric flux ratio, a high X-ray luminosity, and sources located more than 1.5~mag below the main sequence are flagged with an F, L, and M, respectively (see Sect.~\ref{sec: considering additional properties} for details). The full sample is available at CDS. For our analysis, we adopted the unflagged sources as a training set.
\begin{sidewaystable}
        \caption{Basic properties of our training set for the RASS sources}
        \label{tab: training set table}
        \centering
        \resizebox{\textwidth}{!}{
        \begin{tabular}{llllllllllllll}
                \hline \hline
                RASS\_NAME & match\_ID & sigma\_r & sep & p\_geo & p\_ij\_geo & match\_RA & match\_Dec & Fx & HR & G & BP\_RP & plx & flag \\
                &  & arcsec & arcsec &  &  & deg & deg & erg/s/cm2 & mag & mag & mas &  \\
                \hline
                2RXS J213553.8+874703 & 2305328923104539904 & 7.19 & 5.26 & 0.9258 & 0.9258 & 324.008726 & 87.783806 & 6.76e-13 & -0.22 & 8.26 & 0.62 & 11.24 &  \\
                2RXS J041003.1+863735 & 574895911036090752 & 6.03 & 2.72 & 0.9489 & 0.9489 & 62.500553 & 86.626365 & 1.29e-12 & 0.42 & 5.74 & 0.54 & 23.73 &  \\
                2RXS J075556.2+832310 & 1148885991992198016 & 5.64 & 1.43 & 0.9081 & 0.9081 & 118.981225 & 83.386253 & 1.46e-12 & -0.17 & 11.74 & 2.89 & 76.39 &  \\
                2RXS J103104.0+823331 & 1146337324038514176 & 5.24 & 2.12 & 0.9269 & 0.9153 & 157.771328 & 82.558601 & 2.87e-12 & -0.07 & 5.14 & 0.55 & 43.44 &  \\
                2RXS J151801.1+835134 & 1724059113100593920 & 5.19 & 5.85 & 0.9042 & 0.9042 & 229.489581 & 83.859478 & 1.68e-12 & 0.05 & 7.63 & 0.48 & 8.49 &  \\
                2RXS J164557.5+820213 & 1710727122296357376 & 3.90 & 1.24 & 0.9653 & 0.9653 & 251.492329 & 82.037252 & 1.02e-11 & 0.28 & 3.97 & 1.09 & 9.86 &  \\
                2RXS J154017.1+815504 & 1721404028741550848 & 4.20 & 3.99 & 0.9367 & 0.9367 & 235.066581 & 81.917138 & 5.36e-12 & 0.39 & 18.12 & 1.67 & 0.78 & FL \\
                2RXS J221322.5+844540 & 2300493958160639232 & 4.45 & 5.21 & 0.9406 & 0.9406 & 333.332524 & 84.760212 & 6.24e-12 & 0.05 & 9.32 & 1.09 & 16.32 &  \\
                2RXS J220855.9+841349 & 2300460624917976576 & 6.77 & 3.76 & 0.9210 & 0.9210 & 332.236085 & 84.229511 & 5.31e-13 & 0.17 & 11.01 & 1.81 & 20.33 &  \\
                2RXS J234056.5+825218 & 2287506148856660992 & 4.84 & 3.06 & 0.9463 & 0.9417 & 355.242126 & 82.872083 & 1.87e-12 & -0.22 & 8.02 & 1.30 & 50.51 &  \\
                2RXS J215821.0+825218 & 2299477768899190400 & 3.82 & 2.96 & 0.9661 & 0.9660 & 329.585091 & 82.871164 & 2.09e-11 & -0.05 & 7.31 & 0.91 & 25.05 &  \\
                \hline
        \end{tabular}
        }
\tablefoot{The full table is available at CDS}
\end{sidewaystable}

\newpage
\section{Comparison with NWAY}
\label{sec: comparison NWAY}
\begin{figure*}[t]
\centering
\includegraphics[width=0.89\textwidth]{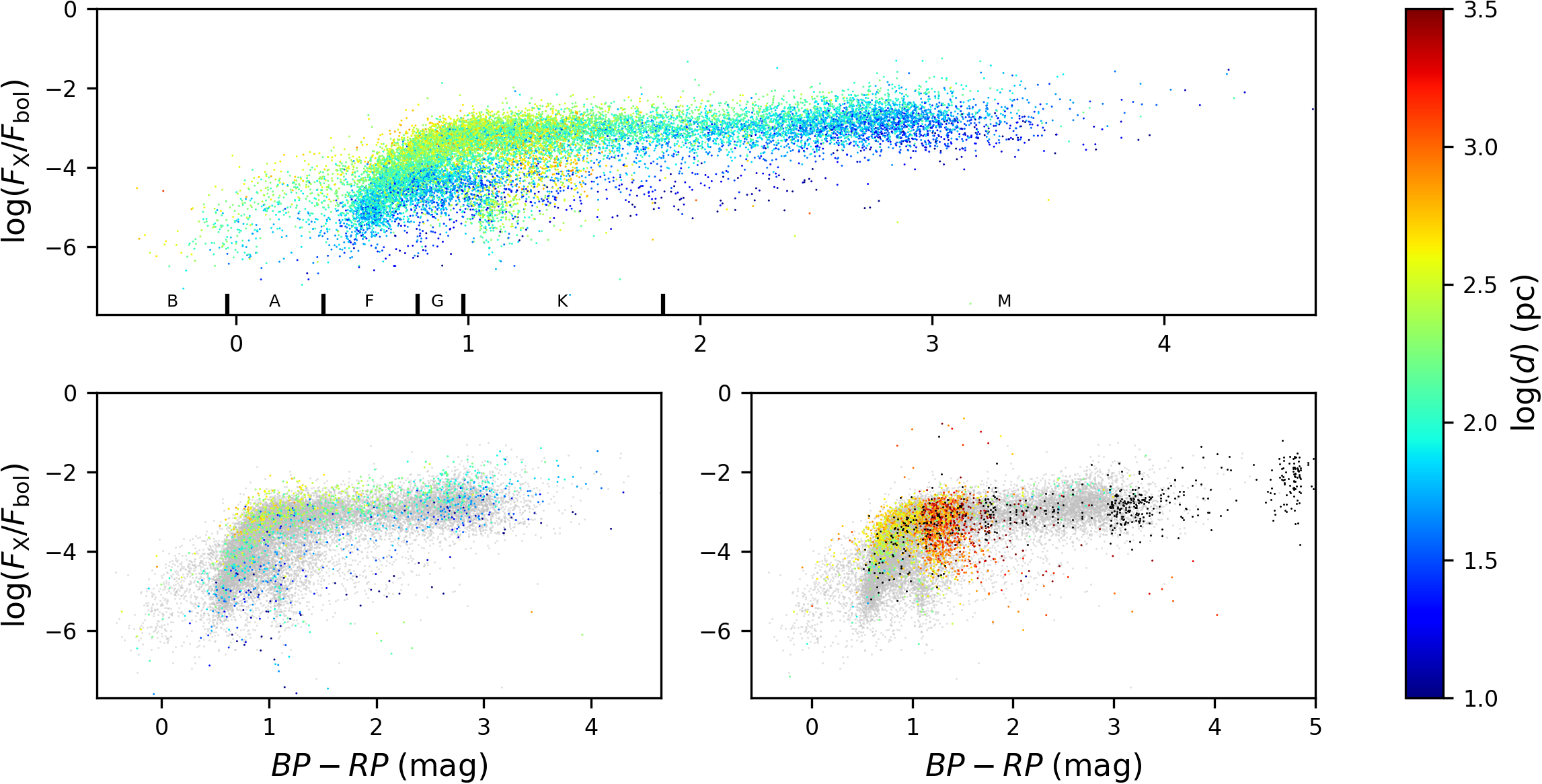}
\caption{Comparison of the X-ray to bolometric flux ratio as a function of the $BP-RP$ color for the stellar RASS sources with a consistent and contradicting NWAY identification. The color scales with the distance of the counterparts, and matches without parallaxes in \textit{Gaia} EDR3 are shown in black. The top, left, and right panels show the distribution for the consistently identified sources, the sources identified by our method but not by NWAY, and the NWAY associations that we did not identify, respectively. The gray dots show the distribution of the top panel as a comparison. 
        \label{fig:comparison NWAY F_X F_bol}}
\end{figure*}

An identification of all RASS sources with Galactic latitudes in excess of 15$^{\circ}$
with the \textit{AllWISE} catalog is presented by \citet{salvato18}, who used NWAY, a Bayesian algorithm for identifying multiwavelength counterparts adopting geometric parameters (e.g., angular separation, positional uncertainty, and counterpart density) and \textit{AllWISE} colors and magnitudes. 
NWAY provides a reliable identification (\texttt{p\_any} $>0.5$) for about 59~\% of the RASS sources with about 5~\% of them being expected to be random associations. 
The goal of \citet{salvato18} is to identify different source types in the RASS catalog but with a special emphasis on AGN, and therefore they excluded RASS sources within the Galactic plane ($|b|<15^\circ$) and with separations smaller than $6^\circ$ and $3^\circ$ to the Large and Small Magellanic Clouds, respectively, to avoid source confusion in regions with high counterpart densities. 

The catalog used by \citet{salvato18} has 90\,850 sources in common with our main catalog, and we identify 18\,739 of them as stellar. 
\citet{salvato18} also provide a relation to discriminate between stars and AGN, according to this criterion 19\,679 of their best counterparts are stars. 
We compared the NWAY identifications with our results by considering \textit{AllWISE} and \textit{Gaia} EDR3 counterparts within 3.5~arcsec to be associated with the same source. 
We find that 17\,284 (92~\%) of the RASS sources were identified with the same counterpart by both methods, but it is important to note that 783 of them do not pass the stellar criterion from \citet{salvato18}. The contradicting identifications are either associated by NWAY to an alternative stellar (499) or nonstellar (573) counterpart or the association is not reliable (\texttt{p\_any} $<0.5$) in NWAY (383). However, for 293 sources, our best identification is given by NWAY as the second best counterpart. There are either two stellar counterparts (175) or a possible stellar and AGN identification (118); in both cases, the RASS source is likely a superposition of two X-ray emitters. On the other hand, 2703 (15.6~\%) stellar counterparts are associated with a RASS source by NWAY, but not by our procedure; 785 sources that we missed were identified by NWAY with a counterpart that is not in our list of stellar candidates. A crossmatch with the SIMBAD database \citep{SIMBAD-database} reveals that some of these sources are indeed stellar sources that are not part of our stellar candidate list, mostly because of a missing parallax measurement in \textit{Gaia} EDR3 though about 47~\% of the sources are classified as extragalactic objects in SIMBAD.

In Fig.~\ref{fig:comparison NWAY F_X F_bol} we compare the X-ray to bolometric flux ratios of the RASS sources consistently identified by both methods with the flux ratios of sources that have contradicting identifications in NWAY and our method. For the counterpart only identified by NWAY without \textit{Gaia} magnitude, we estimated the G-band flux from \textit{2MASS} magnitudes provided in the \textit{AllWISE} catalog applying the conversion of \citet{pec13}. We note that many of the sources with different identifications have $p_\mathrm{stellar}$ values near the cutoff and, hence, are not strictly excluded or confirmed as stellar sources by our method. Some of our stellar identifications not confirmed by NWAY are located at higher activity levels than the main distribution but, overall, the identifications seem to be reasonable. A few of the sources only identified by NWAY as stellar are located far above the saturation limit and some have large distances of more than 1~kpc; we therefore argue that for many of these sources, the X-ray emission is unlikely to be produced by a stellar corona. Some of the NWAY identifications without a Gaia counterpart have very red colors, most of them are classified as extragalactic objects in the SIMBAD database, and for these sources, the conversion between 2MASS and Gaia colors is not valid.

In summary, we conclude that despite NWAY being focused on identifications of extragalactic X-ray sources, the stellar identifications agree reasonably well with our results. The apparently higher number of identifications missed by our procedure is mainly caused by various incompletenesses that we expect to be removed in future \textit{Gaia} releases. 
The stellar selection criterion by \citet{salvato18} produces somewhat different classifications so that about 4~\% of our stellar identifications are missed as stellar by \citet{salvato18}, and, further,
some clearly noncoronal source types are considered stellar.  Due to the fact that \citet{salvato18} considered only X-ray sources above Galactic latitudes of 15$^{th}$ (and outside the
Magellanic Clouds), about a third of the stellar X-ray sky, that is nearly 10\,000 stellar sources, is missed by NWAY.
\end{appendix}

\end{document}